\documentclass[11pt]{article}

\usepackage{amssymb,amsthm}
\usepackage{undertilde}
\usepackage[totalwidth=17cm,totalheight=24cm]{geometry}
\newcommand{\C}{{\mathbb C}}
\newcommand{\R}{{\mathbb R}}

\newcommand{\im}{{\rm i }}

\newcommand\be{\begin{eqnarray}}
\newcommand\ee{\end{eqnarray}}

\begin{document}

\title{Spontaneous symmetry breaking and gravity}
\author{Kirill Krasnov\\ \\ \it{School of Mathematical Sciences, University of Nottingham}\\ \it{University Park, Nottingham, NG7 2RD, UK} \\ and \\ 
\it{Max Planck Institute for Gravitational Physics (Albert Einstein Institute)} \\ \it{Am M\"uhlenberg 1, 14476 Golm, Germany}}

\date{December 2011}
\maketitle

\begin{abstract}\noindent Gravity is usually considered to be irrelevant as far as the physics of elementary particles is concerned and, in particular, in the context of the spontaneous symmetry breaking (SSB) mechanism. We describe a version of the SSB mechanism in which gravity plays a direct role. We work in the context of diffeomorphism invariant gauge theories, which exist for any non-abelian gauge group $G$, and which have second order in derivatives field equations. We show that any (non-trivial) vacuum solution of such a theory gives rise to an embedding of the group ${\rm SU}(2)$ into $G$, and thus breaks $G$ down to ${\rm SU}(2)$ times its centralizer in $G$. The components of the connection charged under ${\rm SU}(2)$ can then be seen to describe gravitons, with the ${\rm SU}(2)$ itself playing the role of the chiral half of the Lorentz group. Components charged under the centralizer describe the usual Yang-Mills gauge bosons. The remaining components describe massive particles. This breaking of symmetry explains (in the context of models considered) how gravity and Yang-Mills can come from a single underlying theory while being so different in the physics they describe. Further, varying the vacuum solution, and thus the embedding of ${\rm SU}(2)$ into $G$, one can break the Yang-Mills gauge group as desired, with massless gauge bosons of one vacuum acquiring mass in another. There is no Higgs field in our version of the SSB mechanism, the only variable is a connection field. Instead of the symmetry breaking by a dedicated Higgs field pointing in some direction in the field space, our theories break the symmetry by choosing how the group of "internal" gauge rotations of gravity (the chiral half of the Lorentz group) sits inside the full gauge group. 
\end{abstract}

\section{Introduction}

The spontaneous symmetry breaking (SSB) scenario plays the fundamental role in the Standard Model (SM) of elementary particles. It assumes the existence of the Higgs field charged under the gauge group to be broken, and transforming as a scalar under the group of isometries of the Minkowski spacetime (the Poincare group). The (renormalizable quartic) potential for this field is then designed so that the field takes a non-trivial vacuum expectation value (VEV), thus breaking the symmetry. The components of the scalar field in the direction of the broken symmetries (would be Goldstone bosons) become the longitudinal components of the gauge bosons, while the remaining component(s) of the Higgs become the massive Higgs particle(s). 

The SM Higgs field couples to fermions via Yukawa couplings, and its non-trivial VEV gives rise to (most of) the fermion masses. Only the (still hypothetical) sterile neutrino sector with its Majorana mass terms does not depend on this mechanism for its mass generation. Thus, the Higgs field and the spontaneous symmetry breaking mechanism for which it is responsible is the essential ingredient in generating the masses of all observed elementary particles. It is of fundamental importance to confirm whether this Standard Model picture is correct and it is therefore not surprising that the search for the Higgs particle is the principal aim of the ongoing expensive, involving large collaborations LHC experiment. A vast amount of the collision data has already been processed, with so far no conclusive outcome. The data analyzed are still compatible with no Higgs being present at all, with however a small excess of events around 125 Gev observed by both major LHC detectors \cite{Higgs}. The data to be collected next year are expected to settle the question of existence or non-existence of the Standard Model Higgs boson. 

It is usually assumed that gravity is of negligible importance for all these phenomena, and can be safely ignored at the energy scales of the accelerator experiments. Still, the fact that it is only the Higgs that introduces the mass scale into an otherwise scalefree Standard Model, and thus tells the SM particles how they should gravitate means that the two are not completely unrelated. At the same time, it is hard to see how gravity can be relevant at the electroweak symmetry breaking energy scales, for it seems to be very weak there. 

While gravity seems to be irrelevant for the physics of elementary particles, it is also very different from all other forces that are carried by gauge fields, at least in its usual metric-based form. The mystery of why the scale where gravity becomes important for quantum fields (Planck scale $10^{19}$ GeV) and the scale where the electroweak symmetry breaking happens (100 GeV) are so different may be not unrelated to the mystery why gravity is so different from the rest of the forces. 

In \cite{Krasnov:2011pp} it was shown that general relativity can be reformulated as particular diffeomorphism invariant ${\rm SU}(2)$ gauge theory. The relevant theory is not Yang-Mills, for the later needs a metric for its definition. Still, it turns out to be possible to write diffeomorphism and gauge invariant Lagrangians for a connection over a 4-dimensional spacetime without using any metric. The arising Lagrangians are non-linear, but lead to second order in derivatives field equations. When the gauge group is taken to be ${\rm SU}(2)$, any of these diffeomorphism invariant gauge theories can be shown to describe interacting gravitons, see \cite{Krasnov:2011up}, and is thus a gravity theory. 

We shall review the construction of diffeomorphism invariant gauge theories in the main text. For now, we would like to ask the reader to take it for granted that such theories are possible, and that for the gauge group $G={\rm SU}(2)$ they describe gravity. Even without going into details of their construction, it turns out to be possible to understand the possible "vacua" solutions of these theories in full generality. Thus, we shall see that any (non-trivial, see below) vacuum solution breaks the full gauge group $G$ of the theory down to an ${\rm SU}(2)$ subgroup, embedded in some way into $G$, plus the subgroup in $G$ that commutes with this ${\rm SU}(2)$ (this may be empty). We shall see that the ${\rm SU}(2)$ components of the full connection then describe gravity, while the components charged under the centralizer of ${\rm SU}(2)$ in $G$ describe the usual Yang-Mills fields. This is a completely general mechanism, independent of details of the diffeomoprhism invariant $G$ gauge theory Lagrangian one starts from. This simple, but completely natural in the context of diffeomorphism invariant gauge theories symmetry breaking mechanism explains why gravity and Yang-Mills fields, while coming from the same underlying theory, can manifest themselves so differently. We will describe the mechanism in more details below.

A natural question that then arises is if the same mechanism can say anything new about the "usual" spontaneous symmetry breaking in the context of gauge theories. The answer to this question is in affirmative, and the underlying mechanism is again very simple and natural. As we shall explain below, the components of the full connection of our theory that are charged under the part of the gauge group that does not commute with the gravitational ${\rm SU}(2)$ give rise to certain massive fields. The following simple scenario then becomes possible. Consider one vacuum where our mechanism breaks $G$ down to ${\rm SU}(2)$ times some smaller subgroup $K\subset G$. As we have said, this will describe $K$ Yang-Mills fields plus gravity (plus certain massive fields for the components of the connection in the directions that do not commute with the gravitational ${\rm SU}(2)$). However, other vacua solutions are also possible, with a different embedding of the gravitational ${\rm SU}(2)$ into $G$. In particular, it is always possible to embed ${\rm SU}(2)$ in such way that it has no non-trivial centralizer, and thus all of the Yang-Mills symmetry group $K$ of the first vacuum is broken. In this latter case all components of the connection apart from those in the "gravitational" ${\rm SU}(2)$ will describe massive fields. Thus, we have a scenario in which by changing the vacuum of the theory one can change the spectrum of the arising excitations from gravity plus massless gauge bosons for some gauge group $K$ (plus massive fields whose nature will be clarified below) to gravity plus a set of massive fields. Intermediate scenarios, in which only some of the symmetries in $K$ are broken, are also possible by choosing the vacuum appropriately. Thus, we have a single mechanism which both explains why gravity is different from the rest of the forces, and that at the same time can be used to break the usual Yang-Mills gauge group as desired. The very availability of such a mechanism, whose details are to be described below, suggests that the spontaneous symmetry breaking in the particle physics and gravity are not so unrelated as they seem to be in the Standard Model. 

We now explain the main ingredient of our mechanism - the construction of vacua of a diffeomorphism invariant gauge theory. This is simple, and can be described in quite non-technical terms already here. Since this construction is so central for everything that follows, and since it is independent of any details of the diffeomorphism gauge theory Lagrangians that we will use later, we decided to explain it already in the Introduction.

The dynamical field of our gauge invariant gauge theories is a connection $A^I, I=1,\ldots,n, n={\rm dim}({\mathfrak g})$ on a 4-dimensional spacetime manifold $M$. Here $\mathfrak g$ is the Lie algebra of $G$, and we use the form notations, so that the connection is a Lie-algebra valued one-form (we always work with trivial bundles here). We would like to describe connections that have a high degree of symmetry and that can therefore be used as the "vacua" of the theory. It is of course always possible to take the zero connection as a "vacuum", but when we describe the construction of the Lagrangians below, we shall see that this $A=0$ configuration is very singular (it is similar to the zero metric configuration in Einstein's GR). Thus, we would like a non-trivial (i.e. different from zero) vacuum connection, which at the same time has enough symmetries to deserve to be referred to as a vacuum solution. The first requirement is that we would like our connection to be the same at all spatial points, i.e. be homogeneous. Thus, we require that the exists a foliation of the spacetime by hypersurfaces such that the connection is the same along every hypersurface (but can possibly change from one such hypersurface to another). This is formalized by introducing some $\eta$ and $x^i$ coordinates, such as $\eta=const$ is a hypersurface of homegeneity, and $x^i$ are some coordinates on these hypersurfaces. Our homogeneous connection can then be written in components as
\be\label{A-backgr}
A^I = a^I_i dx^i + b^I d\eta,
\ee
where $a^I_i, b^I$ are functions of the time coordinate $\eta$ only, as is required by the homogeneity. We now impose yet another requirement that the connection is isotropic, i.e. that at any point with coordinates $x^i_0$ (on any $\eta=const$ hypersurface) we can perform a rotation with $x^i_0$ as the center, and the connection will not change. In other words, we require the connection to be spherically-symmetric with respect to any point $x^i_0$ as a center. It is then convenient to choose the $x^i$ coordinates to be such that $\sum_i (x^i-x^i_0)^2=const$ is a sphere of symmetry with the center $x^i_0$. In other words, we have the usual $\R^3$ coordinates on each spatial hypersurface, and the group of rotations ${\rm SO}(3)$ acts to rotate around a center $x_0^i$ by multiplying $x^i-x^i_0$ with an orthogonal matrix ${\cal O}\in {\rm SO}(3)$. The one-forms $dx^i$ then change as $dx^i\to {\cal O}^i_j dx^j$, and we require the connection $A^I$ to remain invariant under this transformation, i.e. we require that there exists a $G$ gauge transformation that offsets the above transformation of the one-forms $dx^i$. Thus, we require that for any ${\cal O}^j_i \in {\rm SO}(3)$ there exists $g\in G$:
\be
({\rm ad}_g a_i)^I = a^I_j {\cal O}^j_i.
\ee
To understand the meaning of this equation it is convenient to rewrite it in the infinitesimal form. Then we have that for any $\omega_i \in {\mathfrak so}(3)$ there exists $X^I(\omega) \in {\mathfrak g}: f^{I}{}_{JK} X^J(\omega) a^K_i = \epsilon_i{}^{jk} \omega_j a_k^I$, where $\epsilon^{ijk}$ are the usual ${\mathfrak so}(3)$ structure constants and $f^I{}_{JK}$ are those of $\mathfrak g$. We now contract this equation with some other element $u^i\in {\mathfrak so}(3)$. We get: $(a[u,\omega])^I = [X(\omega),au]^I$, where $(au)^I:=a^I_i u^i$. The left-hand-side here is $u,\omega$ anti-symmetric, which means that the right-hans-side should also be. This means that $X(\omega)\sim a\omega$ and that $(a[u,\omega])^I\sim [au,a\omega]^I$. What this equation says is that $a_i^I$ intertwines the action of ${\mathfrak so}(3)$ with that of $\mathfrak g$. Therefore, $a_i^I$ is an embedding of the Lie algebra ${\mathfrak so}(3)$ into $\mathfrak g$
\be\label{embedding}
a_i^I: {\mathfrak so}(3) \to {\mathfrak g}.
\ee
As such, any embedding $a^I_i$ necessarily "breaks" the original gauge symmetry $G$ down to the group of rotations ${\rm SO}(3)$, which is realized as ${\rm SO}(3)\times a({\rm SO}(3))$, i.e. as acting on both indices of $a_i^I$, so that $a_i^I$ is unchanged, times the centralizer of $a({\rm SO}(3))$ in $G$. Note that, depending on the "vacuum" $a_i^I$, i.e. the embedding (\ref{embedding}) chosen, the centralizer $C_a$ of $a({\rm SO}(3))$ in $G$ may be empty (this is what happens in the so-called principal embedding, see below). In this latter case the gauge group is broken completely, down to a single copy of ${\rm SO}(3)$. 

As it will become clear below when we study the Lagrangians arising by linearizing a general diffeomorphism invariant gauge theory around (\ref{A-backgr}), the connection components $a^I_i$ play the role of (a multiple of) the usual Pauli matrices that intertwine the action of the rotation group in space with that of the Lorenz group in the internal space (e.g. on fermions). Thus, our background connection breaks the $G$ symmetry of the theory down to a smaller group (always containing ${\rm SO}(3)$) by essentially selecting how the Pauli matrices sit inside the full gauge group $G$. 

To start with, there is no metric in our theory, the dynamics is described as that of the connection field only. However, as we shall see below, the background connection (\ref{A-backgr}) does introduce a certain metric into the game, which turns out to be that of a constant curvature de Sitter space. Thus, the background (\ref{A-backgr}) admits the usual metric description. To see what are the excitations around the background (\ref{A-backgr}) we linearize a general diffeomorphism invariant $G$ gauge theory Lagrangian around (\ref{A-backgr}). We shall then see that the components of the (linearized) connection in the directions of $a({\mathfrak so}(3))$ in $\mathfrak g$ describe gravitons, while the components in the direction of the centralizer $C_a$ of $a({\mathfrak so}(3))$ in $\mathfrak g$ describe the Yang-Mills fields. The remaining directions in the Lie algebra $\mathfrak g$, i.e. those that do not commute with $a({\mathfrak so}(3))$, will be found to describe certain massive fields that typically transform non-trivially both under $a({\mathfrak so}(3))$ and $C_a$. In other words, they are typically particles of non-zero spin, also charged with respect to the Yang-Mills gauge group. We shall then see that by making the embedding $a$ of ${\mathfrak so}(3)$ to be more and more non-trivial in $\mathfrak g$, we can break more and more of the Yang-Mills symmetries, with the fields that used to be massless gauge bosons in one vacuum becoming massive gauge bosons in another. This is, very briefly, the version of the symmetry breaking mechanism in the context of diffeomorphism invariant gauge theories. 

We would like to finish the Introduction by giving some historical credits. A (preliminary) study of gravity/gauge theory unification of the type considered in this paper was carried out in \cite{TorresGomez:2009gs} and \cite{TorresGomez:2010cd}. However, both works used a formalism close to that of the Plebanski formulation \cite{Plebanski:1977zz} of general relativity. Earlier work on similar ideas, but using the Hamiltonian formalism was reported in e.g. \cite{Chakraborty:1994vx}. Another attempt at gravity-Yang-Mills unification that however uses the full Lorentz group ${\rm SO}(1,3)$ rather than its chiral half is described in \cite{Smolin:2007rx}. One of the aims of the present work is to carry out the analysis using an arguably simpler "pure connection" formalism, where the only variable that the action depends on is a connection field. The simplicity of the "pure connection" formalism to be used will allow to develop a much more complete understanding of the spectrum of the arising particles than was achieved in \cite{TorresGomez:2009gs}. In particular, it is the pure connection formulation that naturally suggests the interpretation of different possible vacua of the theory in terms of a symmetry breaking mechanism. As far as we are aware, the SSB mechanism described here is new. 

Finally, a note of caution is in order. No claim is made here that a realistic unification of gravity with all the fields present in the Standard Model of elementary particles is achieved. What we are after in this paper is a {\it mechanism} that introduces gravity into the story of symmetry breaking, not a realistic model. However, we cannot resist to give a set of examples that come very close to the usual Standard Model symmetry breaking pattern, or even coincide with it. Thus, one example closest to reality that we give is that of ${\rm SU}(6)$ theory that, when broken in a particular way (principal embedding into ${\rm SU}(5)$), gives rise to gravity plus electromagnetism plus the massive spin one $W^\pm$ and $Z$ bosons (plus other more exotic higher spin massive particles). If desired, this can be crossed with the ${\rm SU}(3)$ gauge group to describe the full bosonic content of the Standard Model (without, however, the Higgs boson). Still, our diffeomorphism invariant gauge theories working with ordinary "bosonic" gauge groups $G$ can only describe the bosonic particles. As such, no truly realistic scenario can be possible in this framework, for it necessarily must involve fermions. While it is possible that more realistic scenarios with fermions are possible in the case of $G$ being a super-group, we do not consider such generalizations in the present paper. We reiterate that the main aim of this paper is to exhibit a general mechanism that introduces gravity into the subject of symmetry breaking rather than a realistic model. Work on scenarios possible in this context is to a large extent left to the future. 

The paper is organized as follows. We describe the class of theories that we work with in the next section. Here we also give more details on the "vacuum" solutions and, in particular, verify that the connections described in the Introduction satisfy the field equations. We also show how our background connections give rise to a certain metric, so that the linearized theories we obtain all live in the usual metric background. Section \ref{sec:lin} obtains the linearization of the general Lagrangian around a given background, and also discusses the issue of gauge invariance in the presence of mass terms. We then analyze each sector -- gravity, Yang-Mills and massive -- in turn in sections \ref{sec:grav}, \ref{sec:ym} and \ref{sec:massive}. Examples of our symmetry breaking scenario are treated in Section \ref{sec:sb}. We finish with a discussion.

We warn the readers that some of the analysis presented below is not straightforward, especially in the case of the massive sector. Because of this, we advise to start with Section \ref{sec:sb}, which is self-contained, and treats many examples that illustrate quite well how our symmetry breaking mechanism works in practice. The preceding sections can then be read to understand how our claims about the interpretation of three different sectors of the theory are substantiated.

\section{The theory}
\label{sec:theory}

\subsection{Action principle}

The class of theories that we study in this paper is as follows. Let $A^I$ be the gauge field one-form. The spacetime index is suppressed, $I=1,\ldots, n$ is a Lie algebra index, and $n={\rm dim}(G)$. The gauge group $G$ is an arbitrary (complex at this stage, reality conditions are to be imposed below) Lie group. We shall assume it to be simple (for simplicity of the presentation), but semi-simple direct products of simple groups are also easy to consider, as was done e.g. in \cite{TorresGomez:2010cd} in the context of gravity plus electromagnetism unification. It would be interesting to understand what more general Lie algebras given by semi-direct products give rise to, but we shall not consider this problem in the present paper. The curvature two-form is given by $F^I=dA^I + (1/2)[A,A]^I$. Let $f:{\mathfrak g}\otimes_s {\mathfrak g}\to \C$ be a scalar-valued function on the second symmetric power of the Lie algebra $\mathfrak g$ of $G$ with the following properties: (i) $f$ is gauge invariant $f({\rm ad}_g X)=f(X), \forall g\in G, \forall X\in {\mathfrak g}\otimes_s {\mathfrak g}$; (ii) $f$ is homogeneous function of order one $f(\alpha X)=\alpha f(X), \forall \alpha\in \C^*, \forall X\in {\mathfrak g}\otimes_s {\mathfrak g}$. It is then not hard to see that such a function can be applied to the wedge product $F^I\wedge F^J$ of the curvature two-form with itself, with the result being a well-defined 4-form. This form can then be integrated to produce an action. Thus, the theories we consider are defined by the following action principle
\be\label{action}
S[A]=\im \int f(F\wedge F),
\ee
where $f$ is a function with properties as above (choice of $f$ defines the theory), and the factor of $\im=\sqrt{-1}$ is introduced for future convenience. It has to do with the fact that the reality conditions we will later impose are those corresponding to the Lorentzian signature.

It is worth noting that there are no dimensionful coupling constants involved in the definition of the theory (\ref{action}). Indeed, we take the connection field $A^I$ to have the natural mass dimension one, so that the curvature has the mass dimension two. Due to the homogeneity of $f$, the Lagrangian then has mass dimension four as required. Thus, there are only dimensionless couplings in (\ref{action}). At the same time it should be noted that an infinite number of dimensionless couplings needs to be given to specify a theory from the class (\ref{action}), as it involves specifying a function (of typically many arguments). 

\subsection{Field equations}

The Euler-Lagrange equations that result from (\ref{action}) are as follows:
\be\label{feqs}
d_A \left( \frac{\partial f}{\partial X^{IJ}} F^J \right)=0,
\ee
where, if desired, the factor of $F^J$ can be taken outside of the brackets on which the covariant derivative $d_A$ acts, in view of the Bianchi identity $d_A F^I=0$. The matrix $X^{IJ}$ in (\ref{feqs}) can be defined as $F^I\wedge F^J=({\rm vol}) X^{IJ}$ where $({\rm vol})$ is an arbitrary volume form. The matrix $X^{IJ}$ so defined then depends on the choice of $({\rm vol})$, but the matrix of partial derivatives of $f$ present in (\ref{feqs}) is independent of this ambiguity. 
  
It is not hard to see that the field equations (\ref{feqs}) are second-order in derivatives non-liear partial differential equations for the connection components. It can be argued that the class (\ref{action}) is the most general class of diffeomorphism invariant gauge theories that leads to not higher than second order field equations. 

It can also be shown, see \cite{TorresGomez:2009gs} for an argument to this effect in the context of two-form formulations of the same class of theories, that for a general $f$ the theory (\ref{action}) is a dynamically non-trivial theory with $2n-4$ propagating degrees of freedom (DOF). However, when $f(X)= f_{top}(X)\equiv{\rm Tr}(X)$ the theory becomes topological, with no propagating DOF. 

\subsection{Gravity}

The simplest non-trivial subclass of theories (\ref{action}) arises when $G={\rm SU}(2)$. Indeed, in the simpler case $G={\rm U}(1)$ the only possibility is $f=f_{top}$ and no non-trivial theories exist. Thus, if we are to remain in the simple $G$ context, we are forced to consider non-abelian gauge groups. It can then be shown, see \cite{Krasnov:2011pp}, that Einstein's theory of general relativity can be describe in this language and corresponds to
\be\label{f-GR}
f_{GR}(X) = \frac{1}{16\pi G\Lambda} \left( {\rm Tr} \sqrt{X} \right)^2,
\ee
where $G$ is the Newton's constant, and $\Lambda$ is the cosmological constant. Other choices of $f$ lead to different from GR gravitational theories describing, as GR, just two propagating polarizations of the graviton. In other words, choices of $f$ different from (\ref{f-GR}) (and from $f_{top}$) correspond to different from GR interacting theories of massless spin 2 particles, see \cite{Krasnov:2011up}. We will see how the ${\rm SU}(2)$ diffeomorphism invariant gauge theories manage to describe gravitons below, when we linearize the general action (\ref{action}) around a "vacuum" solution. 

\subsection{Vacua}

One way to see that the theories (\ref{action}) for $G={\rm SU}(2)$ describe interacting massless spin 2 particles is to select an appropriate background and expand the theory around it. The usual graviton excitations of the linearized theory can then be identified, see \cite{Krasnov:2011up}. This is also the strategy we shall follow in the present paper, now for an arbitrary gauge group. In particular, we shall see that different choices of the background (or vacuum of the theory) will lead to different symmetry breaking patterns. 

We now describe the set of "vacua" for (\ref{action}) in more details. In the Introduction we have seen that a general homogeneous isotropic connection is given by (\ref{A-backgr}), with the functions $a^I_i, b^I$ being those of the time coordinate $\eta$ only, and $a^I_i$ defining an embedding of ${\mathfrak so}(3)$ into $\mathfrak g$. We now note that we can always use a time-dependent gauge-transformation to set the $d\eta$ part of the connection (\ref{A-backgr}) to zero. Thus, we are led to consider the following "vacuum" configurations:
\be\label{A*}
A^I = \frac{1}{\im} a^I_i dx^i, \qquad a^I_i: {\mathfrak so}(3)\to {\mathfrak g},
\ee
where we have introduced a factor of $1/\im$ (as compared to (\ref{A-backgr})) for future convenience, and where $a^I_i$ is an embedding of ${\mathfrak so}(3)$ into $\mathfrak g$. Different such embeddings (up to conjugacy inside $G$) are all classified, and for groups $G$ of interest to us here will be described below. For now it is useful to know that for a fixed $G$ there is only a finite number of inequivalent embeddings of ${\mathfrak so}(3)$ into $\mathfrak g$.

Let us now compute the curvature of (\ref{A*}). For this we note that the fact that $a^I_i$ is an embedding of ${\mathfrak so}(3)$ into $\mathfrak g$ means that $f^I{}_{JK} a^J_j a^K_k \sim a^I_i \epsilon^i{}_{jk}$. Let us now represent $a^I_i$ as a constant embedding $e^I_i$ normalized so that
\be\label{e-norm}
f^I{}_{JK} e^J_j e^K_k = e^I_i \epsilon^i{}_{jk},
\ee
times a function $a(\eta)$ of the time coordinate:
\be
a^I_i = a e^I_i.
\ee
We then get for the curvature
\be
F^I = - a^2 e^I_i \left( \im \frac{a'}{a^2} d\eta \wedge dx^i + \frac{1}{2} \epsilon^{ijk} dx^j \wedge dx^k\right).
\ee
It is now convenient to choose a new time coordinate (which we are completely free to reparameterize up to this point) so that 
\be
\frac{a'}{a^2} d\eta = dt.
\ee
The function $a$ then becomes that of $t$, and it is easy to see that we have
\be
a(t)=\frac{1}{t_0-t}, \qquad t< t_0,
\ee
where $t_0$ is an integration constant. Finally, we can conveniently rewrite the above answer for the curvature as
\be\label{curv*}
F^I = - M^2 e^I_i \Sigma^i,
\ee
where
\be\label{Sigma}
\Sigma^i := c^2 \left( \im dt \wedge dx^i + \frac{1}{2} \epsilon^{ijk} dx^j \wedge dx^k\right),
\ee
and
\be
c(t)=\frac{1}{M(t_0-t)},
\ee
where we have introduced an arbitrary dimensionful parameter $M$ (dimensions of mass) so that the function $c(t)$ is dimensionless, and all the dimension of the curvature is carried by $M^2$. As we shall see below, $M$ will be the only dimensioful quantity entering into our construction. All other dimensionful quantities will be expressed in terms of $M$. We note that the objects (\ref{Sigma}) (or rather their Minkowski spacetime limits obtained by setting $c(t)=1$, see below) are often referred to in the literature as 't Hooft symbols \cite{'tHooft:1976fv}. 

To further clarify the meaning of the dimensionful quantity $M$ that we introduced, we note that the background connection (\ref{A*}) with its curvature form (\ref{curv*}) gives rise to a certain spacetime metric. Indeed, it is not hard to recognize in (\ref{Sigma}) the triple of self-dual two forms 
\be\label{Sigma-general}
\Sigma^i = \im \theta^0\wedge \theta^i + \frac{1}{2} \epsilon^{ijk} \theta^j\wedge \theta^k,
\ee
where $\theta^0,\theta^i$ is a tetrad $ds^2 = -(\theta^0)^2 + \sum_i (\theta^i)^2$ for the de Sitter metric 
\be\label{metric}
ds^2 = c^2(t)( - dt^2 + \sum_i (dx^i)^2 ),
\ee
written in the flat slicing, with $c(t)$ as the conformal factor and $t$ being the conformal time. 

In more geometrical terms, the metric appears as follows. We start with a connection (\ref{A*}) and find that its curvature spans a 3-dimensional subspace in the space of all two-forms on our 4-dimensional manifold. Therefore, it can be shown that there exists a conformal metric with respect to which the curvature two-forms (\ref{curv*}) are self-dual. This metric is a multiple of (\ref{metric}). We then fix a representative from the conformal class so that both (\ref{curv*}), (\ref{Sigma-general}) hold and that the proportionality coefficient $M^2$ in (\ref{curv*}) is a constant. It is clear that the parameter $M$ can be arbitrarily rescaled as desired, provided one also rescales the metric in the same way. One can put it differently by saying that what is canonically defined by the curvature two-forms (\ref{curv*}) is a dimensionful metric, such that if used to compute any length it gives a dimensionless result. This dimensionless length are unambiguously defined by the connection (and its curvature), while the parameter $M$ appears if one desires to express this dimensionless length as $M$ times some dimension $1/M$ length computed using a usual dimensionless metric. All in all, the parameter $M$ is introduced to provide a link to the usual physics happening in a metric background. One should keep in mind, however, that $M$ can be arbitrarily rescaled provided the spacetime metric (and thus all length) are also rescaled correspondingly. 

\subsection{Verification of the field equations}

In this subsection we verify that (\ref{A*}) is indeed a solution of the field equations (\ref{feqs}). We have separated this verification into a separate subsection because it is somewhat lengthy. Readers not interested in seeing the details of this computation can skip directly to the next section.

We start by carefully performing the integration by parts that leads to (\ref{feqs}), to fix the notation. Thus, the first variation of the action (\ref{action}) is proportional to 
\be
\delta S[A] \sim \int \frac{\partial f}{\partial X^{IJ}} F^I \wedge \left( d \delta A^J + f^J{}_{KL} A^K\wedge \delta A^L\right),
\ee
where we have dropped an unimportant at this stage overall constant, and have spelled out the covariant derivative. We now integrate by parts in the first term and read off the field equations:
\be
d\left( \frac{\partial f}{\partial X^{IJ}} F^J\right)  - \frac{\partial f}{\partial X^{KJ}} F^J f^K{}_{LI} A^L \equiv 
d_A \left( \frac{\partial f}{\partial X^{IJ}} F^J\right)=0.
\ee
We can now use the fact that for our specific background connection (\ref{A*}) we have $\Sigma^i\wedge \Sigma^j\sim \delta^{ij}$ and thus:
\be\label{X*}
X^{IJ}\sim e^I_i e^J_j \delta^{ij},
\ee
where $\delta^{ij}$ is the usual ${\rm SO}(3)$ invariant metric in ${\mathfrak so}(3)\sim \R^3$. This is a tensor in ${\mathfrak g}\otimes_s {\mathfrak g}$ that is only non-zero on vectors in the image $e({\mathfrak so}(3))\subset {\mathfrak g}$. In particular, this means that the matrix of first derivatives $\partial f/\partial X^{IJ}$ at the background is constant and thus
\be
d\left( \frac{\partial f}{\partial X^{IJ}} \right) =0.
\ee
This is of course only true for our constant curvature background. We can now use the Bianchi identity $d_A F^I=0$ to write $dF^I=-f^I{}_{JK} A^J F^K$. After some simple relabelling of indices we get the following field equation to verify
\be\label{feq-1}
\left( \frac{\partial f}{\partial X^{IL}} f^I{}_{JK} + \frac{\partial f}{\partial X^{IK}} f^I{}_{JL} \right) A^J \wedge F^K=0.
\ee

We now substitute the expressions for the background connection and curvature. We have:
\be
A^J \wedge F^K = \im a^3 e^J_j e^K_k dx^j\wedge \left( \im dt\wedge dx^k + \frac{1}{2} \epsilon^{klm} dx^k \wedge dx^m\right).
\ee 
We see that there are two terms to worry about. One is proportional to $d^3x$. This part of $A^J\wedge F^K$ is proportional to $X^{JK}$. This means that for this components of the 3-form the first term in (\ref{feq-1}) does not contribute as it contains the anti-symmetric structure constant contracted with the symmetric matrix $X^{JK}$. The second term contributes an expression proportional to
\be
 \frac{\partial f}{\partial X^{IK}} f^I{}_{JL} X^{JK},
 \ee
 but this is zero as directly follows from the gauge invariance of the function $f$. Thus, it remains to consider only the $dt \wedge dx^i\wedge dx^j$ components of the 3-form (\ref{feq-1}). We thus need to verify that the following quantity
 \be\label{feq-2}
 \left( \frac{\partial f}{\partial X^{IL}} f^I{}_{JK} + \frac{\partial f}{\partial X^{IK}} f^I{}_{JL} \right) \epsilon^{jkl} e^J_j e^K_k 
 \ee
 is zero. Multiplying (\ref{e-norm}) with another completely anti-symmetric tensor we get
 \be\label{ident-e1}
 f^I{}_{JK} e^J_j e^K_k \epsilon_i{}^{jk} = 2 e^I_i.
 \ee
 Thus, (\ref{feq-2}) can be rewritten as
 \be
 2  \frac{\partial f}{\partial X^{IL}} e^{Il}  + \frac{\partial f}{\partial X^{IK}} f^I{}_{JL} \epsilon^{jkl} e^J_j e^K_k 
 \ee
We now see that both of the terms contain the matrix of first derivatives projected at least in one of its indices on the ${\mathfrak so}(3)$ direction. However, using the fact that $\partial f/\partial X^{IJ}$ must be invariant under the transformations from the embedded ${\mathfrak so}(3)$ we can write
\be\label{ident-1}
\frac{\partial f}{\partial X^{IJ}} e^{J}_i \sim g_{IJ} e^J_i,
\ee 
where $g_{IJ}$ is the Killing-Cartan metric on $\mathfrak g$. This is to be understood as an equation holding when the matrix of first derivatives is evaluated on (\ref{X*}). This equation is just a statement that the pullback of the "metric" $\partial f/\partial X^{IJ}$ to the Lie algebra ${\mathfrak so}(3)$ must be proportional to the pull-back of the Killing-Cartan metric. The proportionality coefficient here is not hard to determine by multiplying this with another factor of $e^I_i$. Using the homogeneity of $f$ this immediately gives
\be\label{ident-2}
\frac{\partial f}{\partial X^{IJ}} e^{J}_i = \frac{f}{{\rm Tr}(X)} g_{IJ} e^J_i,
\ee 
where ${\rm Tr}(X):= g_{IJ} e^I_i e^J_j \delta^{ij}$. Using this identity it is not hard to see that (\ref{feq-2}) is proportional to
\be
2g_{IL} e^{Il} - g_{IK}  f^I{}_{JL} \epsilon^{kjl} e^K_k e^J_j = 0,
\ee
where the last identity follows from (\ref{ident-e1}) and the cyclic symmetry of the structure constants
\be\label{cyclic}
g_{IK} f^I{}_{JL} = g_{IL} f^I_{KJ},
\ee
the last formula being just the statement that ${\rm Tr}(A[B,C])={\rm Tr}(B[C,A])$. Thus, we see that the field equations for the connection (\ref{A*}) hold.
 
\section{Linearization}
\label{sec:lin}

In this section we linearize the action (\ref{action}) around a given vacuum of the form (\ref{A*}). We shall leave the embedding $a^I_i$ (or $e^I_i$ normalized as in (\ref{e-norm})) unspecified for now, and present a general analysis valid for any embedding. 

\subsection{Variations}

For purposes of computing the variations of the action it is convenient to switch to notations in which the spacetime indices are explicit, and introduce a densitiezed matrix
\be
\tilde{X}^{IJ} =\frac{1}{4} \tilde{\epsilon}^{\mu\nu\rho\sigma} F_{\mu\nu}^I F_{\rho\sigma}^J.
\ee
Here $\tilde{\epsilon}^{\mu\nu\rho\sigma}$ is a completely anti-symmetric density that does not need a metric for its definition. This definition is convenient for we have $F^I\wedge F^J = \tilde{X}^{IJ} d^4 x$. Let us evaluate this for the background connection (\ref{A*}). Using the self-duality of the 2-forms $\Sigma^i_{\mu\nu}$:
\be\label{self-duality}
\frac{1}{2} \epsilon_{\mu\nu}{}^{\rho\sigma} \Sigma^i_{\rho\sigma} = \im \Sigma^i_{\mu\nu},
\ee
where $\epsilon_{\mu\nu\rho\sigma}$ is the volume form for (\ref{metric}) and the indices are raised and lowered with this metric, as well as the following algebra of the 2-forms:
\be
\Sigma^i_\mu{}^\rho \Sigma^j_{\rho\sigma} = - \delta^{ij} g_{\mu\nu} + \epsilon^{ijk} \Sigma^k_{\mu\nu},
\ee
we easily get:
\be
\tilde{X}^{IJ} = 2\im M^4 \sqrt{-g} \, m^{IJ},
\ee
where
\be
m^{IJ}:=e^I_i e^J_j \delta^{ij}.
\ee
The above $\tilde{X}^{IJ}$ is the background value where all the derivatives of the function $f$ need to be evaluated. However, because the function is homogeneous, it is sufficient to know the derivatives evaluated at $m^{IJ}$. Thus, we introduce a rescaled matrix $\tilde{X}^{IJ}$, which we shall denote by  symbols with a hat
\be\label{X-hat}
\hat{X}^{IJ}:= \frac{\tilde{X}^{IJ}}{2\im M^4 \sqrt{-g}}.
\ee
When evaluated on the background we have:
\be\label{X-m}
\hat{X}^{IJ}\,\hat{=}\,m^{IJ},
\ee
where the hat symbol over the equality sign means "equal at the background".

Let us now write down the variations of the (densitiezed) matrix $\tilde{X}^{IJ}$. We have:
\be\nonumber
\delta \tilde{X}^{IJ} = \tilde{\epsilon}^{\mu\nu\rho\sigma} F^{(I}_{\mu\nu} D_{\rho} \delta A_\sigma^{J)}, \\
\label{X-vars}
\delta^2 \tilde{X}^{IJ} = \tilde{\epsilon}^{\mu\nu\rho\sigma}\left( 2D_{\mu} \delta A_\nu^{I} D_{\rho} \delta A_\sigma^{J} + F^{(I}_{\mu\nu} f^{J)}{}_{KL} \delta A^K_\rho \delta A^L_\sigma \right), 
\\ \nonumber
\delta^3 \tilde{X}^{IJ} = 6 \tilde{\epsilon}^{\mu\nu\rho\sigma}D_{\mu} \delta A_\nu^{(I}  f^{J)}{}_{KL} \delta A^K_\rho \delta A^L_\sigma.
\ee
The fourth variation is zero. In these formulas $D_\mu$ is the covariant derivative with respect to the background connection. We have denoted it by $d_A$ in the form notations, but in the index notations it is more convenient to use an upper case letter instead of writing the $A$-subscript. 

Let us now evaluate (\ref{X-vars}) at the background, also rescaling to pass to (\ref{X-hat}). We are using self-duality (\ref{self-duality}) freely, without mentioning it. We get:
\be
\nonumber
\delta \hat{X}^{IJ}\,\hat{=}\, -\frac{1}{M^2} \Sigma^{i\,\mu\nu} e^{(I}_i D_{\mu} \delta A_\nu^{J)}, \\
\label{X-hat-vars}
\delta^2 \hat{X}^{IJ} \,\hat{=}\,\frac{1}{\im M^4} \epsilon^{\mu\nu\rho\sigma} D_{\mu} \delta A_\nu^{I} D_{\rho} \delta A_\sigma^{J} - \frac{1}{M^2} \Sigma^{i\,\mu\nu} e^{(I}_i f^{J)}{}_{KL} \delta A^K_\mu \delta A^L_\nu , 
\\ \nonumber
\delta^3 \hat{X}^{IJ} \,\hat{=}\, \frac{3}{\im M^4} \epsilon^{\mu\nu\rho\sigma}D_{\mu} \delta A_\nu^{(I}  f^{J)}{}_{KL} \delta A^K_\rho \delta A^L_\sigma.
\ee
Note that these formulas, unlike (\ref{X-vars}), explicitly depend on the metric (via $\Sigma^{i\,\mu\nu}$ and $\epsilon^{\mu\nu\rho\sigma}$). Note also that the right-hand-sides are invariant under a simultaneous rescaling of the metric $g_{\mu\nu}\to \Omega^2 g_{\mu\nu}$, provided $M^2\to \Omega^{-2} M^2$. In principle, we have the full group of conformal transformations acting (with $\Omega$ being a function), but we have fixed the gauge partially by requiring $M$ to be a constant. It is useful to keep in mind this availability of (constant) conformal rescalings of the metric, as it gives a useful check of the formulas. 

\subsection{Action evaluated on the background}

Since we want to evaluate this on the background, it is convenient to rewrite the action as a function of $\hat{X}^{IJ}$, and then obtain variations of the action using the formulas (\ref{X-hat}). We can write the action as
\be
S[A]=-2M^4 \int d^4 x\sqrt{-g}\, f(\hat{X}).
\ee
We now compare this evaluated on (\ref{X-m}) with the Einstein-Hilbert action for a constant Ricci curvature metric (our signature conventions are $(-,+,+,+)$):
\be\label{action*}
S_{\rm GR}[g]\,\hat{=}\,  -\frac{\Lambda}{8\pi G} \int d^4 x\sqrt{-g}.
\ee
We thus see that our dimensionful parameter $M^4$ plays the role of a multiple of the energy density of the cosmological constant
\be
M^4 f(m) = \frac{\Lambda}{16\pi G}.
\ee
Of course, there is neither $\Lambda$ nor $G$ in our theories, so this interpretation should be taken as an analogy only. However, it is natural to take the (so far arbitrary) length scale $1/M$ to be such that $1/M^2$ is a multiple of the length scale determined by the cosmological constant $\Lambda$. Indeed, in this case the metric (\ref{metric}) will be what the metric of our Universe seems to look like, and the usual identification of scales becomes possible. So, we take $M^2=\Lambda/3$. This gives
\be\label{f-vac}
f(m) = \frac{9}{16\pi G\Lambda}\sim M_p^2/M^2,
\ee
where we have introduced the Planck mass $M_p$. For the observed value of the cosmological constant (and the known value of the Newton's constant $G$) this is a very large number $f(m)\sim 10^{120}$. Thus, it is clear that if we are to extract from our diffeomorphism invariant gauge theories the usual (observed) physics, then the value $f(m)$ of functions $f$ in the vacuum should be very large (\ref{f-vac}). We leave a discussion of whether this is in any sense natural aside for the moment, as this naturalness problem is of course nothing but the reincarnation of the famous cosmological constant problem. But the fact that very large numbers appear will be important below for analyzing the typical mass scales for our theories. 

\subsection{Second variation}

It is now straightforward to compute variations of the action (\ref{action}). For instance, the second variation is given by
\be\label{sec-var-1}
\delta^2 S[A]=-2M^4 \int d^4x \sqrt{-g} \left(\frac{\partial^2 f}{\partial \hat{X}^{IJ} \partial \hat{X}^{KL}} \delta \hat{X}^{IJ}\delta \hat{X}^{KL} + \frac{\partial f}{\partial \hat{X}^{IJ}} \delta^2 \hat{X}^{IJ} \right).
\ee
We then substitute (\ref{X-hat-vars}) and get
\be
\delta^2 S[A] \,\hat{=}\, -2\int d^4 x\sqrt{-g}\, \left( \frac{\partial^2 f}{\partial \hat{X}^{IJ} \partial \hat{X}^{KL}} 
(\Sigma^{i\,\mu\nu} e^{(I}_i D_{\mu} \delta A_\nu^{J)}) (\Sigma^{k\,\rho\sigma} e^{(K}_k D_{\rho} \delta A_\sigma^{L)}) \right.
\\ \nonumber
\left. - \frac{\partial f}{\partial \hat{X}^{IJ}} \left( \im \epsilon^{\mu\nu\rho\sigma} D_{\mu} \delta A_\nu^{I} D_{\rho} \delta A_\sigma^{J} + M^2 \Sigma^{i\,\mu\nu} e^{(I}_i f^{J)}{}_{KL} \delta A^K_\mu \delta A^L_\nu \right) \right),
\ee
where all the derivatives of $f$ are to be understood as computed on $\hat{X}=m$. As the last step we integrate by parts in the first term in the second line. Our final result for the second variation is
\be\label{lin-action}
\delta^2 S[A] \,\hat{=}\, -2\int d^4 x\sqrt{-g}\, \left( \frac{\partial^2 f}{\partial \hat{X}^{IJ} \partial \hat{X}^{KL}} 
(\Sigma^{i\,\mu\nu} e^{(I}_i D_{\mu} \delta A_\nu^{J)}) (\Sigma^{k\,\rho\sigma} e^{(K}_k D_{\rho} \delta A_\sigma^{L)}) \right.
\\ \nonumber
\left. - M^2 \frac{\partial f}{\partial \hat{X}^{IJ}} \left( \Sigma^{i\,\mu\nu} \delta A^{(I}_\mu f^{J)}_{KL} e^K_i \delta A^L_\nu + \Sigma^{i\,\mu\nu} e^{(I}_i f^{J)}{}_{KL} \delta A^K_\mu \delta A^L_\nu \right) \right).
\ee
If desired, the interactions (cubic terms etc) can be obtained in the same way, but we will refrain from considering them in this work. Our task is now to unravel the dynamics encoded in the above linearized Lagrangian.

\subsection{Gauge symmetry}

Before we proceed with the analysis of (\ref{lin-action}) we would like to see how the gauge symmetries of our theory are realized. This is particularly non-trivial given that we seem to have a "mass" term for the connection in (\ref{lin-action}), so it is important to see that this term is compatible with gauge invariance. The gauge symmetry acts on the connection perturbation as
\be\label{M-gauge}
\delta_\phi (\delta A_\mu^I) = D_\mu \phi^I.
\ee
As we shall soon see, it is essential that the full covariant derivative is kept in the kinetic term in (\ref{lin-action}), for otherwise there is no way that the mass term can be invariant under these gauge transformations. The issue of how Minkowski space limit of (\ref{lin-action}) thus gets entangled with the question of how the gauge symmetry (\ref{M-gauge}) can be fixed. 

Let us consider the transformation of (\ref{lin-action}) under (\ref{M-gauge}). We have
\be
\delta_\phi (\Sigma^{i\,\mu\nu} D_\mu \delta A^I_\nu) = \Sigma^{i\,\mu\nu} D_\mu D_\nu \phi^I =   -2 M^2 f^I{}_{JK} e^J_j \phi^K \delta^{ij},
\ee
where we have used $2 D_{[\mu} D_{\nu]} \phi^I = f^I{}_{JK} F^J_{\mu\nu} \phi^K$, substituted the expression (\ref{curv*}) for the background curvature, and then eliminated the contraction of two $\Sigma$'s. We can now write down the full variation of the linearized Lagrangian under gauge transformations. We write it as (half) of the variation of the Lagrangian in (\ref{lin-action}). We have
\be
\delta_\phi {\cal L}^{(2)} = 4M^2 \frac{\partial^2 f}{\partial \hat{X}^{IJ}\hat{X}^{MN}} e^I_i f^J{}_{KL} e^K_j \phi^L \delta^{ij} e^M_k  (\Sigma^{k\,\mu\nu} D_\mu \delta A_\nu^N) 
\\ \nonumber
+2M^2 \frac{\partial f}{\partial \hat{X}^{IJ}} \left( \Sigma^{i\,\mu\nu} \delta A^{(I}_\mu f^{J)}{}_{KL} e^K_i D_\nu \phi^L + \Sigma^{i\,\mu\nu} e^{(I}_i f^{J)}{}_{KL} \delta A^K_\mu  D_\nu \phi^L \right).
\ee
Integrating by parts in the second line to put the covariant derivative on the connection (no additional terms result as all quantities are constant and we have $D_\mu \Sigma^{i\,\mu\nu}=0$), and doing some relabeling of indices we can put the above expression into the form
\be\label{gauge-inv-1}
\delta_\phi {\cal L}^{(2)} = -4M^2\left( \frac{\partial^2 f}{\partial \hat{X}^{IJ}\hat{X}^{MN}} f^I{}_{KL}  \phi^K e^J_i e^L_j \delta^{ij} + \frac{\partial f}{\partial \hat{X}^{IJ}} f^I{}_{KL} \phi^K \delta^L_{(M} \delta^J_{N)} \right) e^M_k  (\Sigma^{k\,\mu\nu} D_\mu \delta A_\nu^N) .
\ee
However, the expression in the brackets vanishes by gauge invariance of $f$. Indeed, the gauge invariance implies
\be
\frac{\partial f}{\partial \hat{X}^{IJ}} f^I{}_{KL} \phi^k \hat{X}^{LJ}=0.
\ee
Differentiating this once more with respect to $\hat{X}^{MN}$ and then evaluating the result on $\hat{X}^{IJ}=m^{IJ}\equiv e^I_i e^J_j \delta^{ij}$ we get precisely the expression in the brackets in (\ref{gauge-inv-1}). 

We learn from the argument just given that it is in general not possible to replace the covariant derivatives in (\ref{lin-action}) while keeping the mass term, as this is inconsistent. Indeed, the terms that would be neglected via this operation are in general of the same order as the mass term. Only with the covariant derivatives can the Lagrangian (\ref{L-M}) be gauge invariant, while at the same time having a non-trivial mass term. This makes the Minkowski spacetime limit of the theory non-trivial to take. Below we shall see, however, that this is possible, but one will have to be careful. In the next two sections we shall see that the mass term is absent in the ${\mathfrak so}(3)$ sector of the theory (gravity), as well as in the sector that commutes with  ${\mathfrak so}(3)$ (Yang-Mills). For both of these sectors the Minkowski limit will be then trivial to take, simply by replacing the covariant derivatives with the usual ones. The remaining sector will be more non-trivial to analyze.  

\section{Gravitational sector}
\label{sec:grav}

\subsection{Linearized Lagrangian for gravitons}

We will start our analysis of (\ref{lin-action}) by demonstrating that the components of the linearized connection $\delta A^I_\mu$ "charged" under the embedded of ${\rm SO}(3)$ describe gravitons. Thus, we decompose the linearized connection as
\be\label{conn-split}
\delta A^I_\mu = \im e^I_i a^i_\mu + a^I_{\perp\,\mu},
\ee 
where $g_{IJ} a^I_\perp e^J_i =0, \forall i$, and we have introduced a factor of $\im$ in the first term for future convenience. We will later also split $a^I_\perp$ into components in the directions that commute with $e({\mathfrak so}(3))$ and in those that do not commute, the two describing very different physics. However, in this section we concentrate on just the $e({\mathfrak so}(3))$ part, i.e. the first term in (\ref{conn-split}).

Let us start by demonstrating that for the present gravitational sector the last term in (\ref{lin-action}) does not contribute. We use (\ref{e-norm}), as well as (\ref{ident-1}), which we rewrite as
\be
\frac{\partial f}{\partial \hat{X}^{IJ}} e^I_i e^J_j = \frac{\delta_{ij}}{3} f(m),
\ee
where we have used the fact that $g_{IJ} e^I_i e^J_j\sim \delta_{ij}$. This immediately shows that the second line in (\ref{lin-action}) (its gravitational $e({\mathfrak so}(3))$ part) is proportional to
\be
\Sigma^{k\,\mu\nu} a^{i}_\mu \epsilon^{ikl} a^l_\nu + \Sigma^{i\,\mu\nu} \epsilon^{ikl} a^k_\mu a^l_\nu =0.
\ee
Thus, we only need to consider the first line in (\ref{lin-action}), i.e. the kinetic term, for this sector. 

Let us consider the tensor
\be
f^{(2)}_{ij|kl}:= \frac{\partial^2 f}{\partial \hat{X}^{IJ} \partial \hat{X}^{KL}} e^I_i e^J_j e^K_k e^L_l.
 \ee
 It is symmetric under the exchange of pairs $ij$ and $kl$, and $ij$, $kl$ symmetric. It is easy to show that its contraction  e.g. $f^{(2)}_{ij|kl} \delta^{kl}$ is zero. Indeed we have
 \be\label{f2-gr-1}
 f^{(2)}_{ij|kl}\delta^{kl}:= \frac{\partial^2 f}{\partial \hat{X}^{IJ} \partial \hat{X}^{KL}} e^I_i e^J_j m^{KL},
 \ee
 but for any $\hat{X}^{IJ}$ we have
 \be
 \frac{\partial^2 f}{\partial \hat{X}^{IJ} \partial \hat{X}^{KL}} \hat{X}^{KL}=0
 \ee
 as follows from the homogeneity of $f$. This can be obtained by differentiating 
 \be
 \frac{\partial f}{\partial \hat{X}^{IJ}} \hat{X}^{IJ} = f
 \ee
 with respect to $\hat{X}^{KL}$. It thus follows that (\ref{f2-gr-1}) is zero and therefore
 \be\label{f2-gr}
 f^{(2)}_{ij|kl}= - \frac{g_{\rm gr}}{2} P^{(2)}_{ij|kl},
 \ee
 where
 \be
 P^{(2)}_{ij|kl} = \frac{1}{2} \left(\delta_{ik} \delta_{jl} + \delta_{il}\delta_{jk}\right) - \frac{1}{3}\delta_{ij}\delta_{kl}
 \ee 
 is the projector on spin 2 representation in the tensor product of two spin 1 representations of ${\mathfrak so}(3)$. It is this appearance of the spin 2 projector that is ultimately related to the components of the connection under consideration describing gravitons. The parameter $g_{\rm gr}$ introduced in (\ref{f2-gr}) is an arbitrary constant, whose value depends on the function $f$ chosen and on the embedding $e^I_i$ that one considers. Its value is unimportant in the linearized theory because it can always be absorbed into the linearized connection. But it does appear as a coupling constant in the interactions, so it is an important parameter of the theory. The minus sign is introduced in (\ref{f2-gr}) so that it is positive for the GR Lagrangian (\ref{f-GR}). 
 
 We can now write down the linearized Lagrangian for $a^i$. It is convenient to rescale $a^i\to a^i/\sqrt{g_{\rm gr}}$ to give $a^i$ the canonical normalization. We also divide the second variation of the action in (\ref{lin-action}) by two to get the correct linearized Lagrangian. We get:
 \be\label{L-GR}
 {\cal L}_{\rm GR}=-\frac{1}{2} P^{(2)}_{ij|kl} (\Sigma^{i\,\mu\nu} D_\mu a^j_\nu) (\Sigma^{k\,\rho\sigma} D_\rho a^l_\sigma),
 \ee
which is the Lagrangian obtained and studied in \cite{Krasnov:2011up}. It is not of the usual form found in the field theory textbooks. Below we shall see that it is very similar in form to the usual Lagrangian of Yang-Mills theory (modulo a surface term). 

Before we proceed with the Hamiltonian analysis of the above Lagrangian, it is worth discussing its gauge symmetries. There are two sets:
\be
\delta_\phi a^i_\mu = D_\mu \phi^i,\qquad \delta_\xi a^i_\mu = \xi^\alpha \Sigma^i_{\alpha\mu},
\ee
with the first one being the usual gauge symmetry, and the second being the realization of diffeomorphisms in our context. Note that the action of diffeomorphisms is very simple, and the corresponding gauge invariance implies that the Lagrangian (\ref{L-GR}) is simply independent of some components of the connection. We shall explicitly see this in the next subsection. 

\subsection{Hamiltonian analysis of the gravitational sector}

Conceptually the most clear way to understand what the Lagrangian (\ref{L-GR}) describes is to perform its Hamiltonian analysis. We are necessarily brief here, to avoid repetition with \cite{Krasnov:2011up}. As a first step, we shall consider (\ref{L-GR}) on scales much smaller than the radius of curvature of our de Sitter background. This is achieved by replacing the covariant derivative $D_\mu$ by the partial derivative $\partial_\mu$ everywhere, and by replacing $\Sigma^i$ with their Minkowski spacetime version obtained by setting $c(t)=1$. Thus, we study (\ref{L-GR}) as a Lagrangian in Minkowski space.

The Lagrangian is built from the combination $\Sigma^{i\,\mu\nu} \partial_\mu a^j_\nu$, and so we compute it using a 3+1 decomposition. We have:
\be
\Sigma_i{}^{\mu\nu} \partial_\mu a^j_\nu = -\im \partial_t a_i^j + \im \partial_i a_0^j + \epsilon_i{}^{kl} \partial_k a_l^j.
\ee
Because of the projector involved in (\ref{L-GR}) we immediately see that only the spin 2 part of the spatial projection $a_i^j$ of the connection is propagating. Its conjugate momentum is
\be
\pi^{ij} := \frac{\partial {\cal L}_{\rm GR}}{\partial \partial_t a_{ij}} = P^{(2)ij}{}_{kl} \left( \partial_t a^{kl} - \partial^k a_0^l +\im \epsilon^{kmn} \partial_m a_n^l \right).
\ee
Here we raise and lower the indices of $a_i^j$ freely using the metric $\delta_{ij}$. Our convention is that the first index in $a^{ij}$ is the spatial one and the second is internal. Here the first term (after the projector is applied) depends only on the spin 2 part of $a^{ij}$. However, the last term contains contributions from the other parts. So, we decompose
\be\label{a-grav-decomp}
a_{ij}= a^{(2)}_{ij} + \epsilon_{ijk} a^{(1)}_{k} + \delta_{ij} a^{(0)}.
\ee
It is then immediately clear that there is no dependence on the spin 0 part. A simple calculation gives
\be
\pi^{ij} = \partial_t a^{(2)\, ij} + P^{(2)\,ij}{}_{kl} \left( - \partial^k (a_0^l +\im a^{(1)\, l}) + \im \epsilon^k{}_{mn} \partial^m a^{(2)\, nl}\right).
\ee
We now introduce the "magnetic" field 
\be
B^{ij}:= P^{(2)\,ij}{}_{kl} \epsilon^k{}_{mn} \partial^m a^{(2)\, nl}
\ee
and write the Hamiltonian ${\cal H}_{\rm GR}=\pi^{ij} \partial_t a^{(2)}_{ij} - {\cal L}_{\rm GR}$ as
\be\label{H-GR}
{\cal H}_{\rm GR}= \frac{1}{2} (\pi^{ij})^2 -\im \pi^{ij} B_{ij} - (a_0^i +\im a^{(1)\, i})\partial^j \pi_{ij},
\ee
where we have integrated by parts in the last term. We see that the last term gives rise to the usual "Gauss" constraint
\be
\partial^i \pi_{ij}=0,
\ee
which generates the gauge transformations of the connection $a^{(2)}_{ij}$
\be
\delta_\phi a^{(2)}_{ij} = P^{(2)}_{ij}{}^{kl} \partial_k \phi_l,
\ee
where $\phi_i$ is a parameter of the gauge transformation. These transformations can be used to set the gauge
\be
\partial^i a^{(2)}_{ij}=0.
\ee
We note that in this gauge the quantity $\epsilon^{ikl} \partial_k a_l^{(2)\, j}$ is automatically symmetric tracefree transverse, so we can simply write $B^{ij}=\epsilon^{ikl} \partial_k a_l^{(2)\, j}$. Finally, to recognize the usual Hamiltonian for gravitons we rewrite (\ref{H-GR}) (with the Gauss constraint already imposed) as
\be
{\cal H}_{\rm GR}= \frac{1}{2} (\pi^{ij}-\im B^{ij})^2 +\frac{1}{2}(B_{ij})^2.
\ee
It is then clear that if we require $a^{(2)}_{ij}$ and 
\be
\tilde{\pi}^{ij}:=\pi^{ij}-\im B^{ij}
\ee 
to be real we obtain the usual graviton Hamiltonian. It can be put into a complete standard form by integrating by parts in the term $(B^{ij})^2$. We get:
\be
{\cal H}_{\rm GR}= \frac{1}{2} (\tilde{\pi}^{ij})^2 -\frac{1}{2} a^{(2)\, ij} \Delta a^{(2)}_{ij},
\ee
which describes two propagating polarizations (symmetric tracefree transverse tensors) of the graviton.

Our final remark is that $\pi^{ij}\to \tilde{\pi}^{ij}= \pi^{ij}-\im B^{ij}$ is a canonical transformation. Indeed, the quantity $\partial_t a^{(2)}_{ij} B^{ij}= \partial_t a^{(2)}_{ij} \epsilon^i{}_{kl} \partial^k a^{(2)\, lj}$ is a total derivative. 
To summarize, we have reproduced the usual free graviton Hamiltonian starting from (\ref{L-GR}), and so verified that the ${\mathfrak so}(3)$ sector of our theory describes gravitons. For further discussion of issues related to this description the reader is referred to \cite{Krasnov:2011up}.

\section{Yang-Mills sector}
\label{sec:ym}

We now study the sector of the theory which is described by the components of the connection in the direction in $\mathfrak g$ that commute with $e({\mathfrak so}(3))$. Thus, we further split $a^I_\perp$ as
\be
a^I_\perp = e^I_a a^a + e^I_\alpha a^\alpha,
\ee
where $e^I_a$ form a basis in the centralizer of $e({\mathfrak so}(3)$ in $\mathfrak g$ and $e^I_\alpha$ spans the remainder of the Lie algebra. Thus, we have
\be\label{i-a-ort}
f^I{}_{JK} e^J_i e^K_a = 0, \qquad \forall i,a.
\ee
With the centralizer being a Lie algebra itself, we are free to choose the basis $e^I_a$ so that
\be
f^I{}_{JK} e^J_a e^K_b = e^I_c f^c{}_{ab},
\ee
where $f^a{}_{bc}$ are the structure constants of the centralizer. 

We now consider just the $a^a$ components of the connection. As in the case of the gravitational sector we start with the second line in (\ref{lin-action}). Using (\ref{i-a-ort}) we see that the first term does not contribute. Then using (\ref{ident-1}) we see that the second term contains $e^I_i f^I{}_{JK} e^J_a e^K_b$, which is zero in view of the cyclic property (\ref{cyclic}) of the structure constants and our assumption (\ref{i-a-ort}) that the index $a$ is in the centralizer. Thus, as in the case of the gravitational sector we only need to consider the first line in (\ref{lin-action}).

Consider the following tensor
\be
f^{(2)}_{ij|ab} :=  \frac{\partial^2 f}{\partial \hat{X}^{IJ} \partial \hat{X}^{KL}} e^I_i e^K_j e^J_a e^L_b.
\ee
Because $a,b$ are in the direction of the centralizer, this tensor must be invariant under independent ${\rm SO}(3)$ rotations of the indices $ij$, as well as under the action of the centralizer of $e({\mathfrak so}(3))$ in $\mathfrak g$ (which we shall denote by $\mathfrak k$) on $ab$. Thus, the above tensor must be proportional to $\delta_{ij}$ times the invariant metric in $\mathfrak k$ if $\mathfrak k$ is simple, or to a linear combination of invariant metrics in $\mathfrak k$ if there are several such metrics. For simplicity we shall assume that there is a unique (modulo rescalings) invariant metric in $\mathfrak k$, which we shall denote by $g_{ab}$, but we should keep in mind that there can be several such invariant tensors. In the latter case the Lagrangian below would involve several (all compatible with the gauge symmetry) terms. In the simplest case of a single invariant tensor we can write
\be
f^{(2)}_{ij|ab} = \frac{g_{\rm ym}}{2} \delta_{ij} g_{ab},
\ee
where $g_{\rm ym}$ is again an arbitrary constant that depends on the function $f$ and the background. We note that unlike in the gravitational sector case, there is no reason why $g_{\rm ym}$ should be large. In fact, this is what becomes (related to) the coupling constant of the Yang-Mills fields once interactions are taken into account. Realistic theories have this of the order of unity, and so we assume that this is the case here as well. We make no attempt here to explain how a function $f$ with such properties can arise, simply using the function with the required properties. 

We can now write the linearized Lagrangian for $a^a$ components of the connection. As in the case of the gravity sector we rescale $a^a\to a^a/\sqrt{g_{\rm ym}}$ to give the connection its canonical normalization. We divide the second variation by two to get the linearized action, and write the resulting Lagrangian
\be\label{L-YM}
 {\cal L}_{\rm YM} = -\frac{1}{2} \delta_{ij} g_{ab} (\Sigma^{i\,\mu\nu} \partial_\mu a^a_\nu) (\Sigma^{j\,\rho\sigma} \partial_\rho a^b_\sigma),
 \ee
 where we have replaced the covariant derivatives by the ordinary partial ones because we now work with the sector that does not transform under the gravitational ${\rm SO}(3)$, and so fact that the background connection is non-trivial is of no significance here. 

We now take into account the fact that
\be
\delta_{ij} \Sigma^{i\, \mu\nu} \Sigma^{j\,\rho\sigma} = 4 P^{\mu\nu\rho\sigma} \equiv
2g^{\mu[\rho} g^{\sigma]\nu} -\im \epsilon^{\mu\nu\rho\sigma}
\ee
is a multiple of the self-dual projector $P^{\mu\nu\rho\sigma}$. This means that modulo a term that is a total derivative the Lagrangian (\ref{L-YM}) is the usual linearized Yang-Mills Lagrangian
\be
{\cal L}_{\rm YM} = -\frac{1}{4} g_{ab} F^{a\,\mu\nu} F^b_{\mu\nu} + {\rm surf.\,\,\, term},
\ee
where $F^a_{\mu\nu} = 2 \partial_{[\mu} a^a_{\nu]}$ is the linearized curvature. We thus see that, as promised, in the sector described by the centralizer $K$ of ${\rm SO}(3)$ in $G$ one has the usual Yang-Mills theory dynamics.

\section{Massive sector}
\label{sec:massive}

\subsection{The mass term}

As we have seen above, both in the gravity and the Yang-Mills sectors the "mass" term in the second line in (\ref{lin-action}) does not contribute. The aim of this section is to analyze the remaining part $a^\alpha$ of the full connection, which are the components charged under part of the Lie algebra that does not commute with the gravitational ${\rm SO}(3)$. We shall see that the mass term is of importance here, and that this remaining sector describes quite an intricate set of fields. Because these fields are generically massive they play a very important role in the symmetry breaking mechanism described in the Introduction.

We start by analyzing the second line in (\ref{lin-action}), i.e. the mass term. The first term here contains the matrix of first derivatives of $f$ projected on the $\alpha$-part of the Lie algebra. In this case we can again, as in (\ref{ident-1}) say that this must be proportional to the Killing-Cartan metric on the Lie algebra
\be\label{ident-3}
\frac{\partial f}{\partial \hat{X}^{IJ}} e^J_\alpha= \kappa \, g_{IJ} e^J_\alpha,
\ee
where the proportionality coefficient $\kappa$, however, can be completely unrelated to that in (\ref{ident-1}).  Now defining
\be
f_{i\alpha\beta}:= g_{IJ} e^I_i f^J{}_{KL} e^K_\alpha e^L_\beta = g_{IJ} e^I_\alpha f^J{}_{KL} e^K_\beta e^L_i,
\ee
where we have used the cyclic property (\ref{cyclic}) of the structure constants we can write the second line in (\ref{lin-action}) as
\be\label{mass-1}
M^2 \left( \kappa - \frac{f(m)}{{\rm Tr}(m)} \right) f_{i\alpha\beta} \Sigma^{i\,\mu\nu}  a^\alpha_\mu a^\beta_\nu,
\ee
where we have used both (\ref{ident-2}) and (\ref{ident-3}). 

Let us compare the two terms in the brackets in the above expression. We know that for realistic theories the quantity $f(m)\sim M_p^2/M^2$ is very large. There is no a priori reason why the parameter $\kappa$ should be large, as in the case of the Yang-Mills sector there was no reason to assume $g_{\rm ym}$ to be large. So, it is safe to assume that
\be
f(m) \gg \kappa
\ee
and drop the first term in (\ref{mass-1}). Note that $M^2 f(m)\sim M_p^2$ so, in case the relevant kinetic terms are canonically normalized, this is the natural mass scale to be expected to arise. Below we shall verify that this is the case. 

\subsection{Kinetic term}

We now consider the kinetic term, i.e. the first term in (\ref{lin-action}). Again, we consider the relevant projection of the matrix of the second derivatives of the defining function $f$:
\be\label{f2-mass-1}
f^{(2)}_{ij|\alpha\beta}:= \frac{\partial^2 f}{\partial \hat{X}^{IJ} \partial \hat{X}^{KL}} e^I_i e^K_j e^J_\alpha e^L_\beta.
\ee
Now that the $e({\mathfrak so}(3))$ and the $\alpha$ parts of the Lie algebra do not commute there are many different invariant tensors that can arise here. A very convenient way to classify them is to think in terms of how the $\alpha$ part of the Lie algebra splits as a representation of the $e({\mathfrak so}(3))$ Lie algebra. This is of course also the way that different embeddings of ${\mathfrak so}(3)$ are classified, see below. So, in general, the part of the Lie algebra $\mathfrak g$ that does not commute with $e({\mathfrak so}(3))$ realizes a representation of ${\mathfrak so}(3)$ and this splits into irreducible representations characterized by their spin. Thus, in general we can write
\be\label{decomp}
V^\alpha = V^{J_1} \oplus V^{J_2} \oplus \ldots \oplus V^{J_k} ,
\ee
where $V^J$ is the irreducible representation of dimension ${\rm dim}(V^J)=2J+1$ and $J_1,J_2,\ldots, J_k$ are the irreducible representations that appear. We note that there can be more than one copy of a representation of the same spin in the decomposition (\ref{decomp}). This occurs e.g. when the centralizer of ${\mathfrak so}(3)$ in $\mathfrak g$ is non-trivial. Then different copies of some $V^J$ of non-trivial multiplicity get mapped to each other by the action of the centralizer. An example of this will be encountered below with the centralizer being ${\mathfrak u}(1)$. In this case, some representations appearing in (\ref{decomp}) are charged with respect to the ${\mathfrak u}(1)$, and thus appear with multiplicity two (as oppositely charged). 

To disentangle the structure present in (\ref{f2-mass-1}) it is very useful to think about $f^{(2)}_{ij|\alpha\beta}$ as a map from the tensor product $V^1\otimes V^\alpha$ to itself. This map must be invariant under both the action of ${\mathfrak so}(3)$ and the action of the centralizer. The representation $V^1\otimes V^\alpha$ splits as
\be\label{J-decomp}
V^1\otimes V^\alpha= \oplus_{J}  \left( V^{J+1}\oplus V^J \oplus V^{J-1}\right),
\ee
where the sum is taken over the representations that appear in (\ref{decomp}). We thus see that in general the same irreducible representation of ${\mathfrak so}(3)$ appears with multiplicity higher than one, even in case that all representations in (\ref{decomp}) have multiplicity one. We can then conclude that $f^{(2)}_{ij|\alpha\beta}$ will be given by a linear combination of projectors on all representations that appear in (\ref{J-decomp}) with multiplicity one, as well as combinations of invariant maps between different copies of the same representation (in case of multiplicity higher than one). 

Our strategy will now be to consider the simplest case of $V^\alpha$ being given by a single representation $V^J$ for some $J$. This is of course too strong of a simplification, as this only arises just for some embeddings into groups $G$ of small rank. However, this example will be very helpful to understand what happens in general. We will come back to the general situation after this particular example is understood.

Thus, assuming that there is a single representation in (\ref{decomp}), from the fact that $f^{(2)}_{ij|\alpha\beta}$ must be a tensor invariant under the ${\mathfrak so}(3)$ transformations preserved by the background, we can deduce that the general form of $f^{(2)}_{ij|\alpha\beta}$ is that of a linear combination of projectors on the 3 different representations appearing in (\ref{J-decomp}) 
\be\label{f2-mass-2}
f^{(2)}_{ij|\alpha\beta} =\frac{1}{2}\left( \kappa_{J+1} P^{J+1}_{ij|\alpha\beta}+ \kappa_J P^{J}_{ij|\alpha\beta}+\kappa_{J-1} P^{J-1}_{ij|\alpha\beta}\right) ,
\ee
where $P^{J}_{ij|\alpha\beta}$ are the corresponding projectors, and $k_J$ are some arbitrary parameters, again related to the properties of the function $f$ and the vacuum chosen. 

With the assumptions made, we can write down the massive sector Lagrangian  connection as 
\be\label{L-M}
{\cal L}_{\rm M} = - f^{(2)}_{ij|\alpha\beta} (\Sigma^{i\,\mu\nu} D_\mu a^\alpha_\nu) (\Sigma^{j\,\rho\sigma} D_\rho a^\beta_\sigma)+ M^2 \left( \frac{f(m)}{{\rm Tr}(m)} - \kappa \right)  f_{i\alpha\beta} \Sigma^{i\,\mu\nu}  a^\alpha_\mu a^\beta_\nu,
 \ee
 where the subscript $M$ in the Lagrangian refers to the massive sector. We emphasize that the connection variable here takes values in some $V^J$ irreducible representation of ${\mathfrak so}(3)$, where $V^J$ is the representations that we assumed the part of $\mathfrak g$ that does not commute with $e({\mathfrak so}(3))$ transforms as. 
 
Before we proceed with our analysis of (\ref{L-M}), it is very instructive to rewrite the identity that the quantity in brackets in (\ref{gauge-inv-1}) is zero in terms of the decomposition (\ref{f2-mass-2}). For this we need to convert the first term into the form present in (\ref{f2-mass-2}). We introduce the pull back of the metric $g_{IJ}$ on the Lie algebra via $e^I_\alpha$:
\be
g_{\alpha\beta}:= g_{IJ} e^I_\alpha e^J_\beta,
\ee
and assume that it is non-degenerate (on $V^\alpha$) so that there is an inverse $g^{\alpha\beta}$. We can then write
\be
\frac{\partial^2 f}{\partial \hat{X}^{IJ}\hat{X}^{MN}} e^I_i e^M_k e^N_\beta = f^{(2)}_{ik|\alpha\beta} g_{IJ} e^I_\gamma g^{\alpha\gamma},
\ee
which is checked by contracting this formula with $e^J_\alpha$. We can now write the identity in question as
\be
\delta^{ij} g^{\gamma\delta} f_{j\gamma\alpha} \phi^\alpha f^{(2)}_{ik|\delta\beta} + \frac{1}{2}\left(\kappa-\frac{f(m)}{{\rm Tr}(m)}\right) f_{k\beta\alpha}\phi^\alpha=0.
\ee
What this identity says is that the components of the matrix of second derivatives in certain directions are related to the matrix of first derivatives. We would now like to identify which of the components in (\ref{f2-mass-2}) are relevant. It is clear that what happens here is the projection of the matrix of second derivatives $f^{(2)}_{ij|\alpha\beta}$ on the direction $f_{i\alpha\gamma}$, i.e. on the direction of the structure constants. Indeed, we can write
\be
P^J_{ij|\alpha\beta}\sim g^{\gamma\delta} f_{\gamma i\alpha} f_{\delta j\beta}.
\ee
We thus learn from this that the coefficient $\kappa_J$ in (\ref{f2-mass-2}) is of the same order as $f(m)$
\be
\kappa_J \sim f(m) \sim M_p^2/M^2, 
\ee
which is very large. There is no reason to assume that the other parameters $\kappa_{J+1}, \kappa_{J-1}$ are as large as $\kappa_J$, as there is no direct relation for them to the gravitational sector, which is the source of the large numbers in our scenario. Thus, we shall assume that $\kappa_{J+1}, \kappa_{J-1}$ are much smaller than $\kappa_J$. 

\subsection{Summary of the problem}

We can now state the (simplified) problem that we need to solve. We have a Lagrangian for a connection taking values in some $V^J$ irreducible representation of ${\mathfrak so}(3)$. It is a Lagrangian of the general form
\be\label{L-M-1}
{\cal L}_{\rm M}^J = - f^{(2)}_{ij|\alpha\beta} (\Sigma^{i\,\mu\nu} D_\mu a^\alpha_\nu) (\Sigma^{j\,\rho\sigma} D_\rho a^\beta_\sigma)+ M^2 f^{(1)}  f_{i\alpha\beta} \Sigma^{i\,\mu\nu}  a^\alpha_\mu a^\beta_\nu,
 \ee
 where $f^{(1)}$ is a parameter, which is very large $f^{(1)}\sim M_p^2/M^2$. The tensor $f^{(2)}_{ij|\alpha\beta}$ is a general covariant tensor that can be thought as an ${\mathfrak so}(3)$ invariant map from $V^1\otimes V^J$ to $V^1\otimes V^J$, and thus splits into a linear combination of 3 projectors (\ref{f2-mass-2}) onto irreducibles (\ref{J-decomp}). Moreover, the Lagrangian is invariant under the gauge transformations, which is essentially the statement that the coefficient $\kappa_J$ in front of the projector on the $J$ representation in (\ref{f2-mass-2}) is proportional to $f^{(1)}$. With the problem reformulated as above, we can forget about where the Lagrangian (\ref{L-M-1}) came from, and study it as it is. 

\subsection{Hamiltonian analysis for $J=1$}
 
Since for a general representation $J$ the Hamiltonian analysis that leads to the description of the propagating degrees of freedom is rather  involved, we would like to simplify things further by treating an example that can be seen to be generic. Thus, we take $J=1$ and work out the Hamiltonian analysis of this case. Simple arguments then show that the pattern for an arbitrary $J$ is exactly the same. 

We first write down (\ref{L-M-1}) explicitly with all the relevant projectors. We have
\be\label{L-M-2}
{\cal L}^1_{\rm M} = - \frac{1}{2}\left( \kappa_2 P^{(2)}_{ij|kl} + \frac{\kappa_1}{2} \epsilon_{ijl} \epsilon_{ljk} + \frac{\kappa_0}{3} \delta_{ij} \delta_{kl} \right) (\Sigma^{i\,\mu\nu} D_\mu a^j_\nu) (\Sigma^{k\,\rho\sigma} D_\rho a^l_\sigma)+ M^2 f^{(1)}  \epsilon_{ijk} \Sigma^{i\,\mu\nu}  a^j_\mu a^k_\nu,
 \ee
where we have written down the $P^{(1)}$ and $P^{(0)}$ representation projectors explicitly. As we know from the discussion in the previous subsection, the coefficient $f^{(1)}$ in front of the mass term is essentially the coefficient in front of the $P^{(1)}$ projector in the kinetic term. This can be fixed from the condition that the Lagrangian is gauge invariant. We have
\be
\delta_\phi (\Sigma_i^{\mu\nu} D_\mu a_\nu^j) = 2M^2 \epsilon^{ijk} \phi^k.
\ee
Substituting this into the variation of (\ref{L-M-2}), integrating by parts in the second term and equating the result to zero we find $f^{(1)}=\kappa_1$ and so the Lagrangian of interest is
\be\label{L-M-3}
{\cal L}^1_{\rm M} = - \frac{1}{2}\left( \kappa_2 P^{(2)}_{ij|kl} + \frac{\kappa_1}{2} \epsilon_{ijl} \epsilon_{ljk} + \frac{\kappa_0}{3} \delta_{ij} \delta_{kl} \right) (\Sigma^{i\,\mu\nu} D_\mu a^j_\nu) (\Sigma^{k\,\rho\sigma} D_\rho a^l_\sigma)+ \kappa_1  M^2  \epsilon_{ijk} \Sigma^{i\,\mu\nu}  a^j_\mu a^k_\nu,
 \ee
 where $\kappa_2,\kappa_1,\kappa_0$ are arbitrary parameters (related to the derivatives of the defining function), and $\kappa_1\sim M_p^2/M^2$. We also note that, as far as the gauge invariance of the above action is concerned, when taking the Minkowski spacetime limit, we are free to replace the covariant derivatives by the ordinary ones in all of the kinetic terms but the one with the coefficient $\kappa_1$ in front. Indeed, we have seen that only that term contributes non-trivially to the cancellation of the variation of the mass term. However, we shall perform this replacement only at the end of the analysis, to keep the formulas for all the three sectors similar. 

We now have all the ingredients to perform the Hamiltonian analysis of the Lagrangian (\ref{L-M-3}). Let us first do the 3+1 decomposition of the kinetic term. The main building block is as before
\be\label{Sigma-Da}
\Sigma_i^{\mu\nu} D_\mu a_\nu^j = -\im \partial_t a^j_i +\im D_i a_0^j +\epsilon_i{}^{kl} D_k a_l^j.
\ee
Here we have used the fact that the background connection (\ref{A*}) only has the spatial components. The kinetic term in (\ref{L-M}) is a square of this block, with the projector on one of the 3 representations appearing in the decomposition (\ref{J-decomp}) applied. Thus, we see that it is natural to split the spatial projection $a^j_i$ of the connection into the 3 irreducible representations (\ref{J-decomp}) of ${\mathfrak so}(3)$ that it contains
\be\label{a-1-decomp}
a_{ij} = a^{(2)}_{ij} + \frac{1}{2}\epsilon_{ijk} a^{(1)}_k + \frac{1}{3}\delta_{ij} a^{(0)},
\ee
which is basically the same decomposition that we were using in the analysis of the gravitational sector, see (\ref{a-grav-decomp}), except that we introduced convenient for the later normalization factors. We now see that each irreducible component of the spatial connection can be propagating, in the sense that there is a non-zero conjugate momentum given by
\be\nonumber
\pi^{(2)\,ij}:= \frac{\partial {\cal L}^1_{\rm M}}{\partial \partial_t a^{(2)}_{ij}} = \kappa_2 \partial_t a^{(2)\,ij} + \kappa_2 P^{(2)\, ij|kl} \left( - D_k a_{0l} +\im \epsilon_k{}^{mn} D_m a_{nl} \right), 
\\ 
\pi^{(1)\,i}:= \frac{\partial {\cal L}^1_{\rm M}}{\partial \partial_t a^{(1)}_{i}} = \frac{\kappa_1}{2} \partial_t a^{(1)\,i} + \frac{\kappa_1}{2} \epsilon^{ikl} \left( - D_k a_{0l} +\im \epsilon_k{}^{mn} D_m a_{nl} \right), 
\\ \nonumber
\pi^{(0)}:= \frac{\partial {\cal L}^1_{\rm M}}{\partial \partial_t a^{(0)}} = \frac{\kappa_0}{3} \partial_t a^{(0)} + \frac{\kappa_0}{3} \delta^{kl} \left( - D_k a_{0l} +\im \epsilon_k{}^{mn} D_m a_{nl} \right).
\ee
Let us now substitute the decomposition (\ref{a-1-decomp}) into the object $\epsilon_k{}^{mn} \partial_m a_{nl}$ and compute its irreducible components. We get
\be\nonumber
P^{(2)\, ij|kl} \epsilon_k{}^{mn} D_m a_{nl} = P^{(2)\, ij|kl} (\epsilon_k{}^{mn} D_m a^{(2)}_{nl} -\frac{1}{2}D_k a^{(1)}_l),
\\ 
\epsilon_{i}{}^{kl} \epsilon_k{}^{mn} D_m a_{nl} = D^m a_{mi}^{(2)} +\frac{1}{2}\epsilon_{ikl} D_k a_l^{(1)} - \frac{2}{3} D_i a^{(0)},
\\ \nonumber
\delta_{kl} \epsilon_k{}^{mn} D_m a_{nl}= D^m a_m^{(1)}.
\ee
We now compute the time derivatives of the connection components in terms of the momenta and then compute the Hamiltonian. We have
\be\nonumber
\partial_t a^{(2)\,ij}  = \frac{\pi^{(2)\,ij}}{\kappa_2} + P^{(2)\, ij|kl} \left( D_k (a_{0l}+\frac{\im}{2} a_l^{(1)}) -\im \epsilon_k{}^{mn} D_m a^{(2)}_{nl} \right), 
\\ 
\partial_t a^{(1)\,i}=\frac{2\pi^{(1)\,i}}{\kappa_1}+\epsilon^{ikl} D_k (a_{0l}-\frac{\im}{2} a_l^{(1)}) - \im D_l a^{(2)\, li} + \frac{2}{3} \im D^i a^{(0)},
\\ \nonumber
\partial_t a^{(0)} = \frac{3\pi^{(0)}}{\kappa_0}+ D^i (a_{0i} -\im a_i^{(1)}).
\ee
We now have
\be
\Sigma_i^{\mu\nu} D_\mu a_\nu^j= -\im \left( \frac{\pi^{(2)\,ij}}{\kappa_2} +  \epsilon^{ij}{}_k \frac{\pi^{(1)\, k}}{\kappa_1} +  \delta^{ij} \frac{\pi^{(0)}}{\kappa_0} \right),
\ee
and so the kinetic term of the Lagrangian is equal to
\be
\frac{1}{2}\left(\frac{(\pi^{(2)})^2}{\kappa_2} + \frac{2(\pi^{(1)})^2}{\kappa_1}+\frac{3(\pi^{(0)})^2}{\kappa_0}\right).
\ee
On the other hand, the mass term in the Lagrangian (\ref{L-M-3}) computes to 
\be
\kappa_1 M^2 \left(2\im a_{0i} a^{(1)\, i} + \frac{2}{3} (a^{(0)})^2 + \frac{1}{2} (a^{(1)})^2 - (a^{(2)})^2 \right),
\ee
which is a term involving the Lagrange multiplier $a_{0i}$, and a set of mass terms for the components of the connection. 

We can now compute the Hamiltonian 
\be
{\cal H}^1_{\rm M}=\pi^{(2)} \partial_t a^{(2)} + \pi^{(1)} \partial_t a^{(1)}+\pi^{(0)} \partial_t a^{(0)}-{\cal L}^1_{\rm M}.
\ee
We get
\be\label{H-1}
{\cal H}^1_{\rm M}=\frac{1}{2}\left(\frac{(\pi^{(2)})^2}{\kappa_2} + \frac{2(\pi^{(1)})^2}{\kappa_1}+\frac{3(\pi^{(0)})^2}{\kappa_0}\right)+ \kappa_1 M^2 \left(\frac{2}{3} (a^{(0)})^2 + \frac{1}{2} (a^{(1)})^2 - (a^{(2)})^2 \right)
\\ \nonumber
-\im \pi^{(2)\,ij} \left( \epsilon_i{}^{kl} D_k a_{lj}^{(2)}-\frac{1}{2}D_i a_j^{(1)} \right) -\im \pi^{(1)\, i} \left( \frac{1}{2} \epsilon_i{}^{jk} D_j a_k^{(1)} + D^j a_{ij}^{(2)} - \frac{2}{3} D_i a^{(0)}\right) - \im \pi^{(0)} D^i a_i^{(1)}
\\ \nonumber
+ a_{0}^i \left( 2\im \kappa_1 M^2 a_i^{(1)} - D^j \pi_{ij}^{(2)} + \epsilon_i{}^{jk} D_j \pi_k^{(1)}-D_i \pi^{(0)}\right),
\ee
where we integrated by parts in the last, constraint, term. It can be checked by an explicit (but a bit involved) computation that the Hamiltonian commutes with the Gauss constraint in the last line.

\subsection{Gauge-fixing and the propagating modes}

The last line in (\ref{H-1}) is the Gauss constraint. We note that, importantly, it generates the usual gauge transformations of the connection components, as well as simple shifts of the momentum $\pi^{(1)}$ variable
\be
\delta_\phi \pi^{(1)\, i} = -2\im \kappa_1 M^2 \phi^i.
\ee
Thus, we can fix the gauge completely by requiring
\be\label{gauge-fix}
\pi^{(1)\, i}=0.
\ee
In this gauge the Gauss constraint equation then gives the connection component $a^{(1)}$ in terms of the momenta $\pi^{(2)}$ and $\pi^{(0)}$. We have
\be\label{a1}
a^{(1)}_i = \frac{1}{2\im \kappa_1 M^2} \left( D^j \pi_{ij}^{(2)}+D_i \pi^{(0)}\right).
\ee
We thus see that the spin 1 sector is non-propagating, and is completely determined by the spin 2 and spin 0 sectors. We now substitute (\ref{gauge-fix}) and (\ref{a1}) into the Hamiltonian (\ref{H-1}). After some elementary transformations (including integrating by parts) we find that the gauge-fixed Hamiltonian decouples into two Hamiltonians for the spin 2 and spin zero sectors correspondingly. We now take the Minkowski spacetime limit, which consists in simply replacing the covariant derivatives by the usual ones. In taking the limit $M\to 0$ we need to take into account that the quantity $\kappa_1 M^2\sim M_p^2$ stays constant. The spin two Hamiltonian is then
\be\label{H2}
{\cal H}_{\rm M}^{1+1} = \frac{(\pi^{(2)})^2}{2\kappa_2}-  \kappa_1 M^2 (a^{(2)})^2 -\im \pi^{(2)\, ij} \epsilon_i{}^{kl} \partial_k a_{lj}^{(2)} - \frac{3}{8\kappa_1 M^2} (\partial^j \pi_{ij}^{(2)})^2.
\ee
The spin zero Hamiltonian is
\be\label{H0}
{\cal H}_{\rm M}^{1-1}=\frac{3(\pi^{(0)})^2}{2\kappa_0}+\frac{2}{3} \kappa_1 M^2 (a^{(0)})^2 +\frac{3}{8\kappa_1 M^2} (\partial_i\pi^{(0)})^2.
\ee
The reason that we wrote the spin two and zero as $1\pm1$ is that the same pattern of only spins $J\pm 1$ being propagating holds for an arbitrary representation $J$, as we will further comment on below. 

\subsection{Evolution equations}

The above Hamiltonians are almost the final answers for the spin $J=1$ sector, except that the fields they contain are not canonically normalized. Instead of performing rescalings of the position and momenta variables (as well as some field redefinitions) to put the Hamiltonians in the canonical form of a massive particle, we exhibit the interpretation of the above systems by simply deriving the equations of motion for the corresponding components of the connection. This is a bit non-trivial exercise for the spin 2 sector because all 5 of the components of this connection propagate and so the connection is not transverse. It is best done by introducing the operator $\epsilon\partial$ that maps symmetric tracefree tensors into same symmetric tracefree tensors. Thus, we define
\be
(\epsilon\partial a^{(2)})_{ij} := P^{(2)}_{ij}{}^{kl} \epsilon_k{}^{mn}\partial_m a_{nl}^{(2)}.
\ee
One then finds that its square is
\be
((\epsilon\partial)^2 a^{(2)})_{ij} = \frac{3}{2} P^{(2)}_{ij}{}^{kl} \partial_k (\partial^m a_{ml}^{(2)}) - \Delta a_{ij}^{(2)}.
\ee
We see that on transverse tensors (such as those appearing in the case of the gravitational sector studied above), the square of $\epsilon\partial$ is just (minus) the Laplacian. In general there is an extra part that involves the transverse part $\partial^j a^{(2)}_{ij}$ of $a^{(2)}_{ij}$. However, this part is cancelled by exactly the same contribution appearing in the field equations derived using (\ref{H2}). The single evolution equation for all 5 of the components of $a^{(2)}$ is then found to be
\be
\ddot{a}^{(2)} = \frac{2\kappa_1 M^2}{\kappa_2} a^{(2)} + \Delta a^{(2)}.
\ee
We see that all 5 components of the spin two field propagate, and that they are massive particles with the mass
\be\label{m-2}
m_2^2 = -\frac{2\kappa_1 M^2}{\kappa_2}.
\ee
This is of the order of the Planck mass squared if $\kappa_2$ is of order unity. However, there is no reason why $\kappa_2$ cannot be chosen to be large, so that the mass is much smaller than $M_p$. Note that the above formula does not imply that the spin two particle is a tachyon, for the defining function $f$ can be chosen so that $\kappa_2$ is negative. 

A similar evolution equation for the spin zero component is much easier to derive, and one gets
\be
\ddot{a}^{(0)}= - \frac{4\kappa_1 M^2}{\kappa_0} a^{(0)} + \Delta a^{(0)},
\ee
from which we identify the corresponding mass as
\be\label{m-0}
m_0^2 =  \frac{4\kappa_1 M^2}{\kappa_0}.
\ee
Again, the scale of this depends on the parameter $\kappa_0$ that can in principle be large. 

All in all, we find that in the analyzed case of the sector $J=1$ the propagating field content of the system is that of massive spin $J\pm 1=0,2$ fields. The masses are given by (\ref{m-2}), (\ref{m-0}), and can in principle be smaller than the Planck scale, if one chooses the defining function of the theory to have large second derivatives in the corresponding directions. Alternatively, one can, if desired, to arrange the spin zero particle to have mass much smaller than $M_p$ so that it is relevant for the low energy physics, and leave the spin 2 particle to have Planckian mass (and thus effectively invisible). We will discuss all these possibilities below, when we talk about different scenarios arising in our framework.

\subsection{Discussion of the general case}

We now make some comments on the general $J$ situation, still being in the context of an (oversimplified) scenario when the complete part of the Lie algebra that does not commute with ${\mathfrak so}(3)$ is a single copy of some irreducible representation $V^J$. The analysis in that case follows exactly the same logic. It is however more non-trivial to spell out, for one needs explicit formulas for the projectors appearing. It is best done using spinor notations, in which the projectors are easy to write. The outcome of this general analysis is exactly the same, for each spin $J$ sector one finds spins $J\pm 1$ as the propagating massive modes. This can be seen rather easily by noting that the mass term will only couple the Lagrange multiplier $a_0^\alpha$ to the $P^J(a^\alpha_i)$ component of the connection. This means that the Gauss constraint generates (apart from gauge transformations of $a^\alpha_i$) shifts in the momentum variable conjugate to $P^J(a^\alpha_i)$, and that therefore this momentum variable can be gauge-fixed away as in (\ref{gauge-fix}). The connection component $P^J(a^\alpha_i)$ is then found from the Gauss constraint equation. Thus, the complete spin $J$ sector does not propagate, leaving us with only $J\pm 1$ propagating spins. The mass parameters for these two sectors are always of the order of $M_p/\sqrt{\kappa_{J\pm1}}$, and so in principle can be very different from $M_p$. We do not see much value in repeating the analysis for the general $J$, given that the result can be expected already from the analyzed case $J=1$. 

Let us now discuss what happens in a more realistic case when there is more than single irreducible representation appearing in (\ref{decomp}). Let us still stay in the setting when all irreps in (\ref{decomp}) appear with multiplicity one. We then look at the decomposition (\ref{J-decomp}) and see that typically (unless the distance between any two $J$'s in (\ref{decomp}) is $|J_1-J_2|\geq 3$) there will appear non-trivial multiplicities. For example, when there are just two spins in (\ref{decomp}), e.g. $J$ and $J+1$, then tensoring with spin one representation gives spins $J-1, J+2$ with multiplicity one, but $J, J+1$ with multiplicity two. The analog of the Lagrangian  (\ref{L-M-1}) in this case will couple all these fields. However, there will be two Gauss constraints that allow one to eliminate one spin $J$ and one spin $J+1$ field. Indeed, this is just eliminating the components of the connection that are pure gauge. So, in this more non-trivial example one will be left with massive fields of spins $J-1, J, J+1, J+2$ propagating. It is a much more non-trivial exercise to disentangle the propagating fields in this case, and we shall not attempt it in this paper. What is important for us is that it is easy to see which spins will be propagating, and that it can in general be expected that all the spins are massive fields. 

The analysis of the general situation follows the same pattern. One looks at the general decomposition (\ref{J-decomp}) from which one erases all the fields that can be eliminated by gauge transformations. The set of propagating fields is then described as 
\be\label{spins}
 \oplus_{J}  \left( V^{J+1}\oplus V^J \oplus V^{J-1}\right) \backslash \oplus_J V^J =\oplus_{J}  \left( V^{J+1}\oplus V^{J-1}\right),
 \ee
 where in general one and the same spin can appear more than once. All these fields should be expected to be massive, with masses determined by the coupling constants appearing from the matrix of second derivatives of the defining function $f$. The natural scale for all the masses is $M_p$, but if the dimensionless couplings arising are allowed to be large than the masses much lower than $M_p$ can be obtained. The masses of all particles of the same spin related by the action of the centralizer of $e({\mathfrak so}(3))$ are the same. 

We now put together all the elements that we have collected so far, and describe the arising symmetry breaking picture.

\section{Symmetry breaking pattern}
\label{sec:sb}

For readers lost in the intricacies of the analysis above, we summarize the picture arising so far in simple terms. Backgrounds of our theory are classified by embeddings of ${\mathfrak so}(3)\sim{\mathfrak sl}(2)$ into $\mathfrak g$, and thus break the gauge symmetry down to a smaller group, which always contains ${\rm SO}(3)$, but in general also contains the centralizer of this ${\rm SO}(3)$ as embedded in $G$. The components of the linearized connection in the direction of ${\rm SO}(3)$ describe usual gravitons, with their two massless propagating polarizations. The components in the direction of the centralizer $K$ of ${\rm SO}(3)$ describe Yang-Mills gauge bosons (two massless polarizations per each Yang-Mills group generator). The components of the linearized connection in the remaining directions in $\mathfrak g$ are massive fields. The best way to understand what they are is to decompose the part of $\mathfrak g$ that does not commute with ${\mathfrak so}(3)$ into irreducible representations of ${\mathfrak so}(3)$. As we shall explain below, this is also the best way to characterize the embedding of  ${\mathfrak so}(3)$ into $\mathfrak g$ in the first place. Let $\{J\}$ be the set of spins that appear. One then obtains a set of typically non-zero spin fields, all massive. The spins that propagate are $J\pm 1$ of all the spins $J\in\{J\}$ that appear in the decomposition of the part of the Lie algebra $\mathfrak g$ that does not commute with the embedded ${\mathfrak sl}(2)$. 

We now illustrate this pattern on the example of algebras ${\mathfrak g}={\mathfrak su}(N)$. However, we first explain how the embeddings of ${\mathfrak so}(3)\sim{\mathfrak sl}(2)$ into ${\mathfrak sl}(N)$ can be classified. 

\subsection{Embeddings into ${\mathfrak sl}(N)$}

The material in this subsection is completely standard, see e.g. \cite{SR}, section 3.1.4 for a particularly useful description. Inequivalent (modulo conjugation) embeddings of ${\mathfrak so}(3)$ into ${\mathfrak sl}(N)$ are classified according to how the fundamental $N$-dimensional representation of ${\mathfrak sl}(N)$ splits into irreducible representations of ${\mathfrak so}(3)$. This splitting gives rise to a pattern
\be\label{fund-split}
N= d_1 + \ldots + d_k,
\ee
where $d$ are integers not equal to zero -- dimensions of irreducible representations of ${\mathfrak sl}(2)$, and one can always order them so that $d_1 \geq d_2 \geq \ldots \geq d_k$. One is usually not interested in the embedding that sends all of ${\mathfrak so}(3)$ to zero, which is eliminated by requiring $d_1>1$. It can be shown that any pattern in (\ref{fund-split}) is possible and corresponds to a unique (modulo conjugation) embedding of ${\mathfrak so}(3)$ into ${\mathfrak sl}(N)$. The corresponding ${\mathfrak so}(3)$ spins are $J=(d-1)/2$. 

While (\ref{fund-split}) classifies the embeddings, for our purposes it is also necessary to know how the adjoint representation of ${\mathfrak sl}(N)$ splits into irreducibles of ${\mathfrak so}(3)$. A convenient in practice way to do this is to note that
\be\label{g-decomp}
{\mathfrak sl}(N) = {\bf N}\otimes {\bf N} - {\bf 0},
\ee
where zero denotes the trivial representation. One then finds the splitting of ${\mathfrak sl}(N)$ into irreducible representations of ${\mathfrak so}(3)$ by simply tensoring together two sums (\ref{fund-split}) and then subtracting the trivial representation of spin zero. Below we shall see how this procedure works on examples. 

\subsection{${\mathfrak sl}(3)$: Massive photon}

There are, up to conjugation, just two different embeddings of ${\mathfrak sl}(2)$ into ${\mathfrak sl}(3)$. We now consider each one in turn, indicating the arising Yang-Mills gauge group (i.e. the centralizer of the embedded ${\mathfrak sl}(2)$) as the title of the corresponding subsection. For the convenience of the reader we also indicate the massive particles content. The upshot of this subsection is that the massless photon of one vacuum becomes massive in the other. This simplest non-trivial setup of ${\mathfrak sl}(3)$ thus provides the simplest example of our spontaneous symmetry breaking mechanism.

\subsubsection{${\mathfrak k}={\mathfrak u}(1)$, massive electrically charged spin 3/2 particle}

The first embedding to consider is what can be called the trivial (or obvious) embedding where one of the two dots of the Dynkin diagram for ${\mathfrak sl}(3)$ is chosen to be image of the $H$-generator of ${\mathfrak sl}(2)$, and a pair positive-negative roots is chosen to be the $E_\pm$ generators. In more basic terms, in this embedding one takes the upper left $2\times 2$ corner of the $3\times 3$ matrices representing ${\mathfrak sl}(3)$ to be the ${\mathfrak sl}(2)$. Under this embedding, the fundamental 3-dimensional representation of ${\mathfrak sl}(3)$ splits as $3=2+1$. In terms of spins, we get the fundamental spin $1/2$ and trivial spin zero representations of ${\mathfrak sl}(2)$. Thus, to determine the decomposition of the adjoint representation we perform the tensor product as in (\ref{g-decomp}):
\be
{\mathfrak sl}(3) = \left( V^{1/2}\oplus V^0 \right) \otimes \left(V^{1/2} \oplus V^0 \right) - V^0 = V^1 \oplus 2 V^{1/2} \oplus V^0.
\ee
Our notation is that $V^J$ is the space of irreducible representation of ${\mathfrak sl}(2)$ of dimension $2J+1$. The $V^1$ part here is the embedded ${\mathfrak so}(3)$, the $V^0$ part is the one-dimensional centralizer generated by the matrix ${\rm diag}(1,1,-2)$, thus giving us a copy of ${\mathfrak u}(1)$. The two copies of the fundamental representation $V^{1/2}$ then span the part of the Lie algebra that does not commute with ${\mathfrak so}(3)$. 

The field content corresponding to this embedding is as follows. We have gravitons for ${\mathfrak so}(3)$, as well as the Maxwell electrodynamics for ${\mathfrak u}(1)$. The remaining part describes massive particles, for which we have twice $V^{1/2}\otimes V^1-V^{1/2}=V^{3/2}$. Thus, in this case there are two spin $3/2$ particles. It is easy to see that they have opposite electric charge (i.e. transform in the opposite way with respect to ${\mathfrak u}(1)$). This is certainly not very realistic example as we have exotic spin $3/2$ particles present. But these can be made to have masses of the order $M_p$, and thus effectively invisible if one so wishes. Then this embedding describes a simple model for gravity plus electromagnetism unified. 

We note that in spite of bosonic particles of half-integer spin being present, one does not encounter problems of the type that seem to be guaranteed by the spin-statistics theorem. The later says that half-integer spin particles must be described by anti-commuting variables, for otherwise the Hamiltonian is unbounded from below. However, as is well-familiar from the textbook case of particles of spin 1/2, this is a result of using Lagrangians that are first order in derivatives. In contrast, our spin 3/2 particles in the example just considered are described by a second derivative Lagrangian. In fact, as can be seen by an explicit computation given in \cite{TorresGomez:2009gs}, the resulting Hamiltonian (in a particular gauge) is that of 4 charged scalar particles. There is no problem whatsoever with describing these particles using commuting variables, as is the case in our construction. We give more comments on the arising higher spin Lagrangians in the discussion section.

The described embedding was first analyzed in \cite{TorresGomez:2009gs}, using however a different (but equivalent) formulation of the theory in terms of Lie algebra valued two-forms. In this reference it was found that there are 4 charged massive scalars (of equal mass) propagating, but it was not realized that these scalars form the spin $3/2$ representation. The present analysis fills this gap. 

\subsubsection{${\mathfrak k}=\emptyset$, massive Maxwell field (plus spin 3 field)}

Let us now consider the other embedding. This is the so-called principal embedding, in which the fundamental 3-dimensional representation of ${\mathfrak sl}(3)$ is irreducible with respect to the ${\mathfrak sl}(2)$ embedded. Thus, we have for the Lie algebra
\be
{\mathfrak sl}(3) = V^1 \otimes V^1 - V^0 = V^1\oplus V^2.
\ee
We see that in this case the centralizer of ${\mathfrak so}(3)$ is empty, and the ${\mathfrak u}(1)$ symmetry of the previous background is completely broken. The $V^1$ part here still describes gravity, but the other part is more interesting. Instead of massless spin 1 particle and two massive spin $3/2$ charged particles of the previous case, all in all containing $2+ 2\times 4= 10$ propagating modes, we now have massive spin $2\pm 1=3,1$ fields, containing the same $7+3=10$ components. We see the the modes reorganized themselves quite non-trivially in this second background. The massless spin 1 Maxwell field of the previous background absorbed one of the models described by two spin $3/2$ particles and became massive. All remaining 7 modes have now combined into a massive spin $3$ field. This example clearly demonstrates that we have all the ingredients of a Higgs mechanism, in which changing the vacuum of the theory changes the spectrum of the modes, with massless gauge bosons of one vacuum becoming massive in the other. 

We thus see that the simplest case of ${\mathfrak g}={\mathfrak sl}(3)$ already gives an example of a mechanism via which the gauge field can be made massive. Note, however, that there is no Higgs boson (massive spin zero particle) in our setup. This is the principal difference with the usual Higgs scenario that would use a charged scalar field (Higgs field) to break the ${\rm U}(1)$ symmetry, and thus give rise to a spin zero Higgs particle. Instead, of the Higgs boson, our scenario gives rise to an exotic  spin 3 particle. If desired, this can be made very massive and thus effectively invisible, so that one gets a simple realistic example of the spontaneous breaking of the ${\mathfrak u}(1)$ gauge symmetry. 

\subsection{${\mathfrak sl}(4)$}

We now consider a more non-trivial, but also more interesting example of ${\mathfrak sl}(4)$ Lie algebra. We now have 4 different embeddings. Again, we treat them in turn, indicating the arising pattern in the title of each subsection.

\subsubsection{${\mathfrak k}={\mathfrak sl}(2)\times {\mathfrak u}(1)$, massive charged spin 3/2 field}

We start with the simplest "trivial" embedding "in the upper left corner". For this we have $4=2+1+1$, and then the Lie algebra splits as
\be
{\mathfrak sl}(4)=\left( V^{1/2}\oplus V^0 \oplus V^0 \right) \otimes \left( V^{1/2}\oplus V^0 \oplus V^0 \right) - V^0 = V^1 \oplus 4 V^{1/2} \oplus 3 V^0 \oplus V^0,
\ee
where we wrote the centralizer as $3 V^0 \oplus V^0$ to indicate that it is in fact the Lie algebra ${\mathfrak sl}(2)\times {\mathfrak u}(1)$, where the ${\mathfrak sl}(2)$ is realized as the algebra generated by $2\times 2$ matrices in the lower right corner of the $4\times 4$ matrices ${\mathfrak sl}(4)$ and ${\mathfrak u}(1)$ is generated by ${\rm diag}(1,1,-1,-1)$. Thus, the unbroken symmetry of this background is a copy of ${\mathfrak sl}(2)$ describing the gravitational sector, as well as ${\mathfrak sl}(2)\times {\mathfrak u}(1)$ describing the unbroken "electroweak" Yang-Mills theory. The other, more exotic particles in this background are 4 spin $3/2$ particles, which can all be viewed as just a single electrically charged massive spin $3/2$ particle transforming in a fundamental representation of the electroweak ${\mathfrak sl}(2)$, together with its anti-particle. 

Any other background will break the above symmetry, and produce a different pattern of excitations. One can break either ${\mathfrak sl}(2)$ symmetry, leaving the ${\mathfrak u}(1)$ intact, or the opposite, or break all of the symmetries. Let us consider the embedding that breaks the ${\mathfrak u}(1)$ symmetry, but preserves ${\mathfrak sl}(2)$. 

\subsubsection{${\mathfrak k}={\mathfrak sl}(2)$, massive scalar (plus spin 2), both in the adjoint of ${\mathfrak k}$. Unbroken electroweak Georgi-Glashow model}

This is the embedding that splits the fundamental 4-dimensional representation as $4=2+2$. We have for the adjoint
\be
{\mathfrak sl}(4)=\left( V^{1/2}\oplus V^{1/2} \right) \otimes \left( V^{1/2}\oplus V^{1/2} \right) - V^0 = V^1 \oplus 3V^1 \oplus 3 V^0.
\ee
This is the embedding that is realized by putting the generators $\sigma^i$ (Pauli matrices) of ${\mathfrak sl}(2)$ into ${\mathfrak sl}(4)$ as ${\rm diag}(\sigma^i,\sigma^i)$, with the diagonal $2\times 2$ blocks filled with $\sigma^i$ and the off-diagonal $2\times 2$ blocks being zero. The centralizer here is in fact a copy of ${\mathfrak sl}(2)$, generated by $\sigma^i$ now understood as a $4\times 4$ matrix in which every element of the Pauli matrix $\sigma^i$ is replaced by a unit $2\times 2$ matrix. So, the symmetry preserved by this background is the gravitational ${\mathfrak sl}(2)$, as well as another copy of Yang-Mills ${\mathfrak sl}(2)$. The ${\mathfrak u}(1)$ of the previous background is broken, as we promised. 

Let us now consider the massive sector of this background. We have 3 massive spin $1\pm 1=2,0$ fields. In other words, we have a massive scalar that transforms as an adjoint representation of the Yang-Mills ${\mathfrak sl}(2)$, as well as a spin 2 particle that also transforms as the adjoint. If desired, the massive spin 2 particle can be made very massive, thus eliminating it from the picture. This leaves us with gravity plus ${\rm SU}(2)$ Yang-Mills plus a massive scalar transforming in the adjoint of ${\rm SU}(2)$. This is the set up of the electroweak Georgi-Glashow model \cite{Georgi:1972cj} with its Higgs in the adjoint, the precursor of the Weinberg-Salam model of the weak interactions. The non-triviality of our example is in the fact that it is here supplemented by gravity. It is also worth emphasizing that we have obtained a propagating spin zero (scalar) particle, something that is not trivial in the present context of gauge theories that work with one-form fields only. 

\subsubsection{${\mathfrak k}={\mathfrak u}(1)$, massive neutral spin 1 (+ spin 3) particles, charged spin 0 (+ spin 2). Electroweak symmetry breaking with scalar $W^\pm$}

With the previous example giving us a Higgs field transforming as the adjoint representation of the weak ${\mathfrak sl}(2)$ we would expect that breaking of this symmetry gives us two charged massive spin 1 particles, and no massive neutral spin 1, as in the Georgi-Glashow model \cite{Georgi:1972cj}. Instead, something else happens, as we now demonstrate. 

So, we now break the ${\mathfrak sl}(2)$ symmetry, while leaving ${\mathfrak u}(1)$ intact. This is achieved by the principal embedding of the gravitational ${\mathfrak sl}(2)$ into the upper left corner ${\mathfrak sl}(3)$. The fundamental representation splits as $4=3+1$, and we have for the adjoint
\be
{\mathfrak sl}(4)=\left( V^{1}\oplus V^{0} \right) \otimes \left( V^{1}\oplus V^{0} \right) - V^0 = V^2 \oplus V^1\oplus 2V^1 \oplus V^0.
\ee
The first of $V^1$ here is the embedded gravitational ${\mathfrak sl}(2)$, the factor of $V^0$ is the unbroken ${\mathfrak u}(1)$, while the irreps $V^2, 2V^1$ correspond to the massive sector. We now apply our $J\pm 1$ rule and see that we obtain electrically neutral spin 3,1 massive particles from $V^2$, and charged particles of spins 2,0 from $2V^1$. This is almost the realistic breaking of the electroweak symmetry, where we have neutral $Z$, and charged $W^\pm$ bosons, except for the fact that our $W^\pm$ bosons are spin 0, not spin 1 particles as they are in the Standard Model. The other exotic spin 3 and spin 2 particles can be eliminated from the picture by making them massive. 

This example illustrates quite well that the patter of symmetry breaking that we get is very distinct from that in the usual Higgs mechanism. Indeed, we would expect to get no neutral spin one particle, while having two charged massive $W^\pm$ bosons. Instead, we did obtain the neutral massive spin one, but the massive $W^\pm$ bosons are scalars, not spin one particles. Below we will present a more realistic scenario where all the massive gauge bosons of the broken electroweak theory are present. 

\subsubsection{${\mathfrak k}=\emptyset$, massive particles of spins 1,2,3,4}

We now consider the principal embedding that breaks all of the symmetries (apart from the gravitational one). The fundamental representation is irreducible $4=4$, and we have
\be
{\mathfrak sl}(4)=V^{3/2}\otimes V^{3/2} - V^0 = V^3\oplus V^2 \oplus V^1.
\ee
The last factor here is the gravitational ${\mathfrak sl}(2)$, while the rest gives rise to massive particles of spins 1,2,3,4. The spin 1 massive particle here can be thought of as the massive gauge boson arising by breaking the ${\mathfrak u}(1)$ gauge symmetry present in the original background. 

\subsection{${\mathfrak sl}(5)$}

This set of examples is particularly interesting because one finds precisely the content of the unbroken Weinberg-Salam model in one of them, with the electrically charged Higgs in the fundamental of the electroweak ${\mathfrak sl}(2)$. However, breaking the symmetry further no Standard Model symmetry breaking pattern arises, which shows that the mechanism here is indeed different from the usual Higgs mechanism. We shall encounter the Standard Model SB patter in the context of the ${\mathfrak sl}(6)$ setup. 

\subsubsection{${\mathfrak k}={\mathfrak sl}(3)\times {\mathfrak u}(1)$, massive charged spin 3/2 field}

We again start from the simplest possible embedding, also the one that leaves as much as possible of the symmetry unbroken. The fundamental representation splits as $5=2+1+1+1$, and we have
\be
{\mathfrak sl}(5)=\left( V^{1/2}\oplus 3 V^0  \right) \otimes \left( V^{1/2}\oplus 3 V^0 \right) - V^0 = V^1 \oplus 6 V^{1/2} \oplus 9 V^0.
\ee
We thus again see a massive spin 3/2, electrically charged, as well as transforming in the fundamental representation of ${\mathfrak sl}(3)$. 

This is the pattern that generalizes to the trivial embedding into ${\mathfrak sl}(N)$, in that one finds the group ${\mathfrak sl}(N-2)\times {\mathfrak u}(1)$ unbroken, with a single massive electrically charged spin 3/2 field transforming in the fundamental of ${\mathfrak sl}(N-2)$. All other symmetry breaking patters arise by breaking some of these ${\mathfrak sl}(N-2)\times {\mathfrak u}(1)$ symmetries with the help of the spin 3/2 Higgs fields. So, we can, if desired, think about our symmetry breaking mechanism as the usual Higgs mechanism that, however, works not with Higgs scalars, but with Higgs spin 3/2 fields. It is then possible to give these fields a non-trivial VEV without breaking the Lorentz symmetry by considering non-trivial embeddings of ${\mathfrak sl}(2)$ into $\mathfrak g$. 

\subsubsection{${\mathfrak k}={\mathfrak sl}(2)\times {\mathfrak u}(1)$, spin 0 (+spin 2) neutral, in the adjoint of ${\mathfrak sl}(2)$, spin 3/2 charged in the fundamental of ${\mathfrak sl}(2)$}

We now consider the fundamental split as $5=2+2+1$, with the adjoint decomposing as
\be
{\mathfrak sl}(5)=\left( 2 V^{1/2}\oplus V^0  \right) \otimes \left( 2 V^{1/2}\oplus V^0 \right) - V^0 = 4 \left( V^1 \oplus V^0 \right) \oplus 4 V^{1/2}.
\ee
The massive sector here is composed of $3V^1$ and $4V^{1/2}$ representations, which give rise to electrically neutral spin 0,2 fields transforming in the adjoint of ${\mathfrak sl}(2)$, as well as electrically charged fields of spin 3/2 transforming in the fundamental of ${\mathfrak sl}(2)$. This is not a very illuminating example. 

\subsubsection{${\mathfrak k}={\mathfrak sl}(2)\times {\mathfrak u}(1)$, massive electrically charged spin 0 (+spin 2) field in the fundamental of ${\mathfrak sl}(2)$, neutral spin 1 (+spin 3) field. Standard Model Higgs field}

The next example has exactly the same symmetry breaking pattern as the previous one, but much more interesting massive field content. The fundamental splits as $5=3+1+1$, and the adjoint as
\be
{\mathfrak sl}(5)=\left( V^{1}\oplus 2V^0  \right) \otimes \left( V^{1}\oplus 2 V^0 \right) - V^0 = V^2 \oplus V^1 \oplus 4V^1 \oplus 4 V^{0}.
\ee
Thus, the massive sector here is composed of $V^2$ and $4V^1$ representations, which give rise to spin 1,3 massive fields that are neutral with respect to the full unbroken Yang-Mills group ${\mathfrak sl}(2)\times {\mathfrak u}(1)$. The four copies of $V^1$ on the other hand give rise to a charged field of spin 0,2 transforming in the fundamental of ${\mathfrak sl}(2)$. Making all the fields apart from the charged field in the fundamental of ${\mathfrak sl}(2)$ very massive we have the set up of the Weinberg-Salam model of electroweak interactions, before the symmetry breaking. The spin zero field is just the usual Standard Model Higgs. We shall see, however, that it is not possible to obtain the Standard Model symmetry breaking pattern staying in the ${\mathfrak sl}(5)$ setting. It does arise, but one has to take a larger gauge group ${\mathfrak sl}(6)$, as we shall see below. 

\subsubsection{${\mathfrak k}={\mathfrak u}(1)$, neutral spin 0,1,2,3, charged spin 1/2, 3/2, 5/2 massive fields}

Our next two examples break the electroweak ${\mathfrak sl}(2)$ while leaving ${\mathfrak u}(1)$ intact. They are interesting from the point of view of seeing that in spite of the Standard Model Higgs field being present here, we do not get the Standard Model SB pattern. We start with the splitting of the fundamental as $5=3+2$, with the adjoint splitting as
\be
{\mathfrak sl}(5)=\left( V^{1}\oplus V^{1/2}  \right) \otimes \left( V^{1}\oplus V^{1/2} \right) - V^0 = V^2 \oplus V^1 \oplus V^{0} \oplus 2\left( V^{3/2}\oplus V^{1/2}\right) \oplus V^1.
\ee
The massive sector here consists of $V^2, V^1, 2V^{3/2}, 2V^{1/2}$ representations. The first two give rise to electrically neutral spin 0,1,2,3 fields. The second two give rise to electrically charged spin $1/2, 3/2, 5/2$ fields. The most interesting feature of this example is the presence of an electrically neutral scalar. 

\subsubsection{${\mathfrak k}={\mathfrak u}(1)$, neutral spin 1,2,3,4, charged spin 1/2, 5/2 massive fields}

In this example the fundamental splits as $5=4+1$ and the adjoint as
\be
{\mathfrak sl}(5)=\left( V^{3/2}\oplus V^{0}  \right) \otimes \left( V^{3/2}\oplus V^{0} \right) - V^0 = V^3 \oplus V^2 \oplus V^1 \oplus V^{0} \oplus 2 V^{3/2}.
\ee
The massive sector here is made from $V^3, V^2, 2 V^{3/2}$ representations, giving rise to electrically neutral spin 1,2,3,4 fields, as well as electrically charged spin $1/2, 5/2$ fields. Thus, we do get the neutral Z-boson in this example, but our charged $W^\pm$ particles have the wrong spin: $1/2$ instead of 1 as in the Standard Model.

\subsubsection{${\mathfrak k}=\emptyset$, massive particles of spins 1,2,3(twice),4,5}

In this last example all of the symmetries are broken, and the fundamental is an irreducible representation of the embedded gravitational ${\mathfrak sl}(2)$. Thus, $5=5$ and the adjoint splits
\be
{\mathfrak sl}(5)=V^{2} \otimes V^{2} - V^0 = V^4 \oplus V^3 \oplus V^2 \oplus V^1 .
\ee
The massive sector here consists of spin 1, 2, 3 (twice), 4,5 fields. This example generalizes to the principal embedding into ${\mathfrak sl}(N)$, where one gets massive fields of spins 1, 2, 3 (twice), 4 (twice), $\ldots, N-3$ (twice), $N-2$ (twice), $N-1$, $N$.

\subsection{${\mathfrak sl}(6)$. Standard Model symmetry breaking pattern}

Our last example is the setting of ${\mathfrak sl}(6)$ gauge group. We will not treat all of the different 9 possible embeddings, leaving this as an exercise to the reader. We will describe only a single embedding that is interesting because it gives rise precisely to the Standard Model symmetry breaking pattern. This is the embedding in which the fundamental representation splits as $6=5+1$, and so we have for the adjoint
\be
{\mathfrak sl}(6)=\left( V^{2}\oplus V^{0}  \right) \otimes \left( V^{2}\oplus V^{0} \right) - V^0 = V^4\oplus V^3 \oplus V^2 \oplus V^1 \oplus V^{0} \oplus 2 V^{2}.
\ee
The group is broken here down to a copy of ${\mathfrak u}(1)$ (times the gravitational ${\mathfrak sl}(2)$), as is appropriate for the Standard Model electroweak sector after the symmetry breaking. The key ingredient here is 3 copies of the spin two representation, with one of them being neutral, and the other two being electrically charged with the opposite charge. These give rise to spin 1 (plus spin 3) neutral field that can be interpreted as the Z boson, as well charged fields of spin 1 (plus spin 3) that can be interpreted as $W^\pm$ bosons. The spin 3 components here can all be made very massive to eliminate them from the picture. The other content comes from $V^4,V^3$ representations, that give rise to massive neutral spin 2,3,4,5 particles, which can again all be made very massive. Thus, we get an example that includes all the particles of the symmetry broken Standard Model (without the Higgs boson but with exotic massive higher spin particles), as well as gravity. 

\section{Discussion}

We recapitulate the main points of our construction. We described a class of connections that have a high degree of symmetry and that satisfy the field equations of our diffeomorphism invariant gauge theories. The connections are in one-to-one correspondence with embeddings of ${\mathfrak sl}(2)\sim{\mathfrak so}(3)$ into the Lie algebra ${\mathfrak g}$ of the gauge group of the theory. For a given gauge group, there is only a finite number of inequivalent (modulo conjugation) embeddings, and thus a finite number of "vacua". 

We did not yet discuss the question of how the theory chooses dynamically which vacuum to live in. However, it is clear that  the action of the theory takes a different value on each vacuum. This value can be interpreted as the integral of $f(m) M^4\sim \Lambda/G$, where $\Lambda$ is the corresponding energy density. Thus, as in the usual Higgs mechanism we can simply postulate that the vacuum that gives the absolute minimum of $f(m)$ is chosen, where $m^{IJ}=e^I_i e^J_j \delta^{ij}$ is the value of the matrix $X^{IJ}\sim F^I\wedge F^J$ that corresponds to a given embedding $e^I_i$. It is clear that choosing the defining function $f$ of the theory appropriately we can make any of the possible vacua to be the absolute minimum. 

We have also seen that each vacuum solution gives rise to a spacetime metric, and that this spacetime metric is in all cases the constant curvature de Sitter metric (anti de Sitter spacetime is also possible, and corresponds to simply taking the parameter $M$ to be imaginary). The radius of curvature of this de Sitter space $1/M$, or the curvature $M^2$ is completely arbitrary. But, once it is fixed, all distances in the theory are measured in units of $1/M$. This is the only place where a dimensionful quantity enters into the construction, as there is no dimensionful coupling constants in our theories. Thus, all dimensionful quantities in our setup get expressed as multiples of powers of $M$. In other words, the dimensionful parameter $M$ plays the role of the unit of mass (and thus length) for our theories. It is very convenient to fix $M$ to be related to the observed value of the cosmological constant as $\Lambda\sim M^2$, for then the de Sitter metric that we use here becomes close to the metric of our Universe as seems to be currently observed, and the usual identification of distances becomes possible. 

We then linearize the general diffeomorphism invariant gauge theory Lagrangian around a particular vacuum solution. The connection perturbations have three different components: (i) those charged under the image of ${\mathfrak sl}(2)$ in ${\mathfrak g}$, and we refer to these components as those describing the gravitational sector; (ii) those charged under the centralizer of the embedded ${\mathfrak sl}(2)$ in ${\mathfrak g}$, and we refer to these as the Yang-Mills sector; (iii) those charged under the part of the Lie algebra ${\mathfrak g}$ that does not commute with the embedded ${\mathfrak sl}(2)$, and these are referred to as the massive sector. We have then shown that, for a generic defining function $f$, the gravitational sector connection components describe gravitons with their two propagating polarizations, and the Yang-Mills components describe the linearized gauge bosons with the usual Lagrangian. The analysis of the massive sector is more involved, but we have seen that this in general describes massive particles of various spins. The spin content can be determined by looking at how the part of the Lie algebra that does not commute with ${\mathfrak sl}(2)$ splits as a representation of ${\mathfrak sl}(2)$. This is in general a sum over irreducible representations of some spins $J$. The propagating massive fields are those of spins $J\pm 1$. We have then analyzed a large set of examples and showed how the symmetry breaking mechanism works in practice. 

Before discussing the symmetry breaking patterns that are possible, we would like to emphasize that some of the Lagrangians arising from our construction are not standard. This is in particular the case for the gravitational sector, where we describe spin two particles (gravitons) using ${\mathfrak sl}(2)$ valued connections. The same is true for the massive sector, where we have seen that typically higher spin fields arise. However, the arising Lagrangians are non-standard, and in particular, not those that are encountered in the higher spin literature. For instance, we have seen the half-integer spins are possible and quite generic. One might immediately jump to a conclusion that the mechanism described here is sick, as it often leads to bosonic particles of half-integer spin, seemingly violating the spin-statistics theorem. However, one should bear in mind that the theorem is proved for a certain type of Lagrangians. Its textbook case of spin 1/2 particles is for the Dirac Lagrangian that is first order in derivatives. The requirement of positivity of energy is then the main reason for having to use the Fermi statistics for the spin 1/2 fields described by the Dirac Lagrangian. In contrast, the Lagrangians that arise from our construction are always second order in derivatives. After the gauge is fixed one can always view the massive fields arising as a collection of scalar fields, which, however, transform non-trivially under the chiral half of the Lorentz group, forming an irreducible representation of some spin $J$. Because of the second derivative nature of the Lagrangian, the Hamiltonian for our massive fields is just the sum of positive definite Hamiltonians of a set of scalar fields, and so no problems with positive definitness arise. Thus, there is no need to use Fermi statistics for our half-ingeral spin fields, and in any case this would not be possible as they came from a bosonic connection. We appreciate that this aspect of our construction is very non-standard, but it is clear just by inspecting the arising positive definite Hamiltonians that there is no problem at least at the linearized level studied. We will come back to the problem of interactions below. 

Another interesting aspect of our construction is that it gives a theory of only a finite number of higher spin particles (together with gravity and possibly Yang-Mills fields), interacting on a constant curvature background. As is well-known from the higher spin literature, the usual construction requires an infinite number of higher spin fields for a consistent theory. This gives another illustration to the fact that our Lagrangians cannot be the usual higher spin Lagrangians. The characteristic features of the Lagrangians that appear from our construction is that they use {\it one-forms} with values in some representation of the chiral half of the Lorentz group as the basic fields, not spacetime tensors of some rank as in the usual constructions. Our Lagrangians also use the 't Hooft symbols (self-dual two-forms) to contract the spacetime and internal indices. This is the other difference with the usual case where no such objects is used. It would be very interesting to understand if there is any relation between our setup and the usual higher spin theories based on spacetime tensors. We do not attempt this in the present work, however. 

To summarize, once expanded around one of the vacua our theory gives rise to gravitons and Yang-Mills gauge bosons, as well as to a set of massive fields of generally non-zero spin. However, as we have seen from the examples above, the spin zero massive particles are also possible. Our mass generation scenario is rather different from that of the Higgs mechanism, and this is worth discussing in more details. Above we have seen that our linearized gauge theory Lagrangian has a non-trivial mass term and is nevertheless invariant under gauge transformations. This is only possible because we work in a constant curvature background (de Sitter space), and the wedge product of two covariant derivatives gives the background curvature that is of the order $M^2$. Then the term that comes from the variation of the kinetic term of the action cancels the variation of the mass term. This is a familiar mechanism for mass generation in a constant curvature background. In fact, it is well-known that e.g. gravitons in a homogeneous isotropic FRW Universe can be viewed as particles of a time dependent mass. What we have here is very similar to this well-known phenomenon. What is unusual, however, is that the mass term survives even after the limit to zero curvature is taken. The way this happens is as follows. From (\ref{mass-1}) it is clear that the mass term comes with a prefactor $f(m)$ of the defining function evaluated on the background matrix $X^{IJ}\sim m^{IJ}=e^I_i e^J_j \delta^{ij}$.  In fact, it is the matrix of the first derivatives of the function $f$ that appears, but it gets related to the value of the function itself by the homogeneity of $f$. We then know that the value $f(m)$ should be given the interpretation of the vacuum energy density, and thus identified with $\Lambda/(G M^4)\sim M_p^2/M^2$. The overall coefficient in front of the mass term is then of the order $M_p^2$ and does not go to zero when the limit $M\to 0$ is taken. In other words, the matrix of first derivatives of the defining function $f$ has some components (precisely in the gravitational ${\mathfrak sl}(2)$ directions) that blow up when one takes $M\to 0$. This blowing up components get multiplied by $M^2$ in the mass term, and result in a finite non-zero mass even in the Minkowski spacetime limit. 

We would now like to say a few words about the hierarchy problem(s) in our context. Our construction does not say anything new about this, in that there is still no explanation of why so large numbers as $M_p^2/M^2\sim 10^{120}$ or $M^2_{W,Z}/M^2\sim 10^{86}$ appear. But what is very interesting is that in our scenario all these large numbers have becomes a part of the defining function $f$. They are all some components of the matrices of (e.g. second) derivatives of the function $f$ with respect to the matrix $X^{IJ}\sim F^I\wedge F^J$, evaluated on the background $X^{IJ}\sim e^I_i e^J_j \delta^{ij}$. As such, they encode some properties of the function $f$. One can now at least in principle ask the question why functions $f$ with such properties are those relevant at low energies. It is then not impossible that some renormalization group flow argument (assuming that $f$ runs with energy) can explain the form of $f$. Thus, one can envisage that the famous cosmological constant problem and the particle physics hierarchy problem can all be addressed in the current framework. In particular, it is very interesting that it is the cosmological constant that becomes the main (and only) dimensionful parameter of the theory. Thus, in our context the cosmological constant problem is not to explain why the $\Lambda$ is so small, but rather to explain why the Planck mass $M_p$ is so large. In other words, the problem can be reformulated as that of explaining why the strength of the graviton interaction is observed to be $E^2/M_p^2$, where $E$ is the energy involved, rather than the value $E^2/M^2$ that would be more natural from the point of view of this formulation of gravity. And indeed, the graviton interaction strength as predicted by our theory turns out to be of the order $E^2/(g_{\rm gr} M^2)$, which requires $g_{\rm gr}\sim f(m)$ to be of the order $M_p^2/M^2\sim 10^{120}$ to be consistent with the observed extremely weak character of gravity. As we have already mentioned, it is not impossible that some renormalization group flow argument for $f$ (together with the asymptotic safety scenario, see below) can provide an explanation for the large numbers needed, and thus for the weakness of the gravitational interaction.

We finally turn to the particular symmetry breaking models that are provided by our construction. We have seen that in many aspects our SSB scenario behaves like the usual Higgs mechanism. In particular, we have seen that in the context of ${\mathfrak sl}(3)$ gauge group it is possible to give the photon a mass by changing the vacuum, this being the simplest mass generating example, an analog of the Maxwell field coupled to a charged scalar field with the Mexican hat potential. The principal difference, however, is that our scenario does not predict a massive spin zero particle -- Higgs boson -- instead giving rise to more exotic higher spin fields. Further, we have seen that the two simplest setups for the electroweak symmetry breaking, namely the unbroken Georgi-Glashow \cite{Georgi:1972cj} and Weinberg-Salam models can both be realized, the first in the context of ${\mathfrak sl}(4)$ gauge group with the $4=2+2$ embedding, the second in the set up of ${\mathfrak sl}(5)$ and $5=3+1+1$ embedding. However, the patters of symmetry breaking are rather different, and we find the usual $Z,W^\pm$ spin one massive gauge bosons only in the set up of ${\mathfrak sl}(6)$ and the $6=5+1$ embedding. All in all, in spite of many similarities our mechanism of symmetry breaking and gauge field mass generation is distinct from the Higgs mechanism. 

In spite of being able to reproduce the familiar Standard Model electroweak symmetry breaking pattern, we are far from claiming that our theories can provide a realistic particle physics model. Indeed, our theories are bosonic, and can never describe fermions of the Standard Model. On this we note that it is very suggestive that one should try to extend the constructions of this paper to Lie superalgebras, or diffeomorphism invariant supergauge theories. These are not hard to formulate, and it remains to be seen if one can use the Grassmann and spinor-valued one-forms that would arise to describe the usual spin 1/2 fermions of the Standard Model. Work on this is in progress. Before this work is completed it is hard to say whether it is possible to have a realistic particle physics model (including fermions) in the framework of diffeomorphism invariant gauge theories. For this reasons, we are consciously avoiding making any predictions from the purely bosonic models considered here, even though it is very clear that these models point in the direction of massive higher spin particles rather than the spin zero Higgs boson as the yet unobserved content of a symmetry broken theory. 

We would like to conclude this paper with some remarks on the open problems of this approach. One of them, namely a description of fermions in this language, has already been mentioned. Another important question is if the theories described here can be considered as fundamental, or they are just some effective field theoretic models. Our first remarks is that these theories containing gravity are necessarily non-renormalizable in the usual sense, i.e. they contain negative mass dimension coupling constants (in front of higher derivative interactions). In fact, they contain an infinite number of such higher derivative interactions, and an infinite number of the corresponding coupling constants, the later being encoded in the defining function $f$. It has then been conjectured (in \cite{Krasnov:2006du} for the case of ${\rm SU}(2)$ theory describing pure gravity) that the present class of diffeomorphism invariant gauge theories with second order in derivative field equations is closed under renormalization. In other words, the conjecture is that the higher derivative terms that need to be added to the theory in the process of its renormalization are already contained in our Lagrangian. As one possible motivation for this conjecture we mention a very interesting analysis presented in \cite{Anselmi:2002ge}. It concludes that if a theory starts as one with a second-derivative propagator, it will stay this way after all renormalizations and all field redefinitions. Our conjecture is similar to this, but in the context of diffeomorphism invariant gauge theories. 

If the class of theories that we studied in the paper is closed under renormalization, then the renormalization group flow is one in the space of defining functions $f$. As such, it can be realistically studied. In particular, a very important question is if there are any ultra-violet (UV) fixed points of the flow. In case the answer is in the affirmative, one can then invoke the asymptotic safety scenario of Weinberg, see \cite{Weinberg:2009bg} for a recent reference, as a possible mechanism for the UV completion of this class of theories. In fact, one particular point in the theory space, namely $f_{top}={\rm Tr}(F\wedge F)$ gives rise to a topological theory, and thus necessarily a fixed point of the RG flow. It is not impossible that this is the fixed point giving the UV completion of this class of theories. The questions of closedness under renormalization, and then of the nature of the arising RG flow are easiest to address in the context of pure gravity theory corresponding to $G={\rm SU}(2)$, and work on this is in progress. To summarize, it is not impossible that the theories considered in this work, in spite of being non-renormalizable in the usual sense, possess a UV completion in the sense of asymptotic safety and are in this sense fundamental. If all the hopes about the UV completion are realized, and if realistic couplings to fermions are possible in this context, we would obtain a large class of UV complete quantum theories containing all interactions (including gravity), which is the hope that drives the present research program. 

Finally, another very serious issue with the scenario described here is that of the unitarity of the models considered. We have started our discussion with a complex gauge theory, and imposed reality conditions on various fields as is appropriate in order to have positive definite linearized Hamiltonians. This is always possible at the linearized level, given the fact that we have the option of multiplying our fields by an imaginary unit if this is necessary to get a positive definite kinetic term. The mass terms can also all be made positive by choosing the defining function appropriately. However, there is no guarantee that the theory with these linearized reality conditions turns out to be unitary when interactions are switched on. This is a particularly difficult problem in view of the fact that in our description all the fields are chiral in the sense that only the chiral half of the Lorentz group plays the role. Thus, imaginary unit is ubiquitous, and the arising Lagrangians are generally complex. The issue of unitarity thus becomes very non-trivial and it is by no means guaranteed that it will be possible to find a satisfactory solution. The strategy that is presently followed is, however, to first see whether the realistic particle physics content (fermions, massive gauge fields) can be described using this language. In this paper we have made the first step in this directions and have shown that the symmetry breaking and the gauge boson mass generation is possible (and quite natural as we have seen above). If it is found that the Standard Model fermions can be similarly described, it will then be sensible to attempt to address the unitarity problem. Thus, the open problems of unitarity together with those of describing fermions and the UV completion are all left to future work.

\section*{Acknowledgements} I am grateful to Alex Torres-Gomez for collaboration on the initial stages of this project. I would like to thank Stefan Theissen for a useful reference \cite{SR}, Andrea Campoleoni for useful discussions on the embeddings of ${\mathfrak sl}(2)$ into ${\mathfrak sl}(N)$, and Roberto Percacci and Dario Benedetti for a discussion on general aspects of this symmetry breaking scenario. I am also grateful to Roberto Percacci for pointing out to me the reference \cite{Anselmi:2002ge}, and to Illarion Melnikov for asking about implications of the spin-statistics theorem. The author was supported by a fellowship from Alexander von Humboldt foundation, Germany.


\begin{thebibliography}{99}

\bibitem{Higgs} "Update on the search for the Higgs boson by the ATLAS and CMS experiments at CERN", December 2011, available at http://cdsweb.cern.ch/record/1406786

\bibitem{Krasnov:2011pp}
  K.~Krasnov,
  ``Pure Connection Action Principle for General Relativity,''
  Phys.\ Rev.\ Lett.\  {\bf 106}, 251103 (2011).
  [arXiv:1103.4498 [gr-qc]].

\bibitem{Krasnov:2011up}
  K.~Krasnov,
  ``Gravity as a diffeomorphism invariant gauge theory,''
  Phys.\ Rev.\  {\bf D84}, 024034 (2011).
  [arXiv:1101.4788 [hep-th]].
  
\bibitem{TorresGomez:2009gs}
  A.~Torres-Gomez, K.~Krasnov,
  ``Gravity-Yang-Mills-Higgs unification by enlarging the gauge group,''
  Phys.\ Rev.\  {\bf D81}, 085003 (2010).
  [arXiv:0911.3793 [hep-th]].
  
\bibitem{TorresGomez:2010cd}
  A.~Torres-Gomez, K.~Krasnov, C.~Scarinci,
  ``A Unified Theory of Non-Linear Electrodynamics and Gravity,''
  Phys.\ Rev.\  {\bf D83}, 025023 (2011).
  [arXiv:1011.3641 [gr-qc]].
  
\bibitem{Plebanski:1977zz}
  J.~F.~Plebanski,
  ``On the separation of Einsteinian substructures,''
  J.\ Math.\ Phys.\  {\bf 18}, 2511 (1977).
  
\bibitem{Chakraborty:1994vx}
  S.~Chakraborty and P.~Peldan,
  ``Towards a unification of gravity and Yang-Mills theory,''
  Phys.\ Rev.\ Lett.\  {\bf 73}, 1195 (1994)
  [arXiv:gr-qc/9401028].
  
\bibitem{Smolin:2007rx} 
  L.~Smolin,
  ``The Plebanski action extended to a unification of gravity and Yang-Mills theory,''
  Phys.\ Rev.\ D {\bf 80}, 124017 (2009)
  [arXiv:0712.0977 [hep-th]].
  
\bibitem{'tHooft:1976fv}
  G.~'t Hooft,
  ``Computation of the Quantum Effects Due to a Four-Dimensional
  Pseudoparticle,''
  Phys.\ Rev.\  D {\bf 14}, 3432 (1976)
  [Erratum-ibid.\  D {\bf 18}, 2199 (1978)].
  
  \bibitem{SR}
  A.~V.~Razumov and M.~V.~Saveliev, "Lie Algebras, Geometry, and Toda-type Systems", Cambridge lecture notes in Physics No. 8, Cambridge 1997. 
  
\bibitem{Georgi:1972cj} 
  H.~Georgi and S.~L.~Glashow,
  ``Unified weak and electromagnetic interactions without neutral currents,''
  Phys.\ Rev.\ Lett.\  {\bf 28}, 1494 (1972).
  
\bibitem{Krasnov:2006du} 
  K.~Krasnov,
  ``Renormalizable Non-Metric Quantum Gravity?,''
  hep-th/0611182.
  
\bibitem{Anselmi:2002ge} 
  D.~Anselmi,
  ``Absence of higher derivatives in the renormalization of propagators in quantum field theories with infinitely many couplings,''
  Class.\ Quant.\ Grav.\  {\bf 20}, 2355 (2003)
  [hep-th/0212013].
  
\bibitem{Weinberg:2009bg}
  S.~Weinberg,
  ``Effective Field Theory, Past and Future,''
  PoS C {\bf D09}, 001 (2009)
  [arXiv:0908.1964 [hep-th]].
  
\end{thebibliography}
\end{document}